\documentclass[11pt,a4paper]{article}

\usepackage{iftex}
\ifPDFTeX\pdfoutput=1\fi
\usepackage{jheppub}
\usepackage[T1]{fontenc}
\usepackage[utf8]{inputenc}
\usepackage[english]{babel}
\usepackage{xcolor}
\usepackage{slashed}
\usepackage{diagbox}
\usepackage{tabularx}
\usepackage{amsfonts}
\usepackage{multirow}
\usepackage{float}
\usepackage[section]{placeins}
\usepackage{microtype}
\usepackage[colorlinks=true, linkcolor=blue, citecolor=blue, urlcolor=blue]{hyperref}

\hypersetup{hidelinks,hypertexnames=false}

\renewcommand{\thetable}{\Roman{table}}
\makeatletter
\let\JHEPoriginalmakecaption\@makecaption
\long\def\@makecaption#1#2{%
  \def\JHEPtabletype{table}%
  \ifx\@captype\JHEPtabletype
    \vskip\abovecaptionskip
    {\normalsize\noindent\tablename\nobreakspace\thetable. #2\par}%
    \vskip6pt
  \else
    \JHEPoriginalmakecaption{#1}{#2}%
  \fi}
\makeatother
\title{NLO QCD corrections to $W^{+}$-boson decays into doubly heavy baryons}

\author[1,2]{Hong-Hao Ma}
\emailAdd{honghao.ma@unesp.br}
\author[3]{Xu-Chang Zheng,}
\emailAdd{zhengxc@cqu.edu.cn}
\author[1]{Juan-Juan Niu,}
\emailAdd{niujj@gxnu.edu.cn}
\author[3]{Xing-Gang Wu}
\emailAdd{wuxg@cqu.edu.cn}

\affiliation[1]{Department of Physics, Guangxi Normal University,\\Guilin 541004, China}
\affiliation[2]{Instituto de F\'{\i}sica Te\'orica, Universidade Estadual Paulista,\\
Rua Dr. Bento Teobaldo Ferraz, 271- Bloco II, 01140-070 S\~ao Paulo, SP, Brazil}
\affiliation[3]{Department of Physics and Chongqing Key Laboratory for Strongly Coupled Physics, Chongqing University, Chongqing 401331, China}

\date{\today}

\abstract{

The next-to-leading-order (NLO) QCD corrections have been investigated for the production of doubly heavy baryons via the inclusive $W^{+}$-boson decays $W^{+}\to\Xi_{cc}+\bar{c}+\bar{s}+X$ and $W^{+}\to\Xi_{bc}+\bar{b}+\bar{s}+X$ within the nonrelativistic QCD framework. The decay widths, the kinematic distributions, and the uncertainties induced by the heavy-quark masses and the renormalization scale have been analyzed for both the color-antitriplet ($\bar{\mathbf{3}}$) and the color-sextet ($\mathbf{6}$) $S$-wave states. With the central input parameters $m_{c}=1.8~\mathrm{GeV}$ and $m_{b}=5.1~\mathrm{GeV}$, the $K$ factors are $2.27$ for $\Xi_{cc}$ and $2.95$ for $\Xi_{bc}$, i.e., the NLO corrections increase the corresponding LO widths by about $127\%$ and $195\%$, which shows that the NLO corrections are indispensable for quantitative predictions. The individual-channel $K$ factors range from $1.94$ to $5.53$, with the largest one occurring for $\Xi_{bc}$ in the $[{}^{3}S_{1}]_{\mathbf{6}}$ channel. The distributions show that the NLO corrections enhance the low invariant-mass region and shift the baryon energy spectrum toward smaller energy fractions. With $\Gamma_{W}=2.085~\mathrm{GeV}$, the inclusive branching ratios are $8.86\times10^{-5}$ and $3.63\times10^{-6}$ for $\Xi_{cc}$ and $\Xi_{bc}$, respectively, which provide an NLO benchmark for future experimental searches.

}

\begin{document}
\maketitle

\section{Introduction}
\label{sec:intro}

Doubly heavy baryons contain two heavy quarks and one light quark and therefore probe several dynamical scales inside the same hadron. If the relative velocity $v$ of the heavy pair is small, the scales $m_{Q}$, $m_{Q} v$, and $m_{Q} v^2$ may be separated and treated within nonrelativistic effective theories~\cite{Caswell:1985ui, Bodwin:1994jh, Brambilla:2004jw}. For the production of doubly heavy baryons in nonrelativistic QCD (NRQCD), a compact colored $Qc$ ($Q=b$ or $c$) diquark is formed at short distances and subsequently evolves into a baryon~\cite{Ma:2003zk}. Effective-field-theory analyses of the $QQq$ system and heavy-quark--diquark symmetry make the assumptions underlying this description explicit~\cite{Fleming:2005pd, Cohen:2006jg}. We therefore treat heavy-diquark factorization as a phenomenological expansion whose assumptions are stated explicitly.

The direct production of ground-state and excited doubly heavy baryons has been studied at leading order (LO) in hadronic collisions~\cite{Baranov:1995rc, Chang:2006eu, Chang:2006xp, Chang:2007pp, Zhang:2011hi, Chen:2014frw, Chen:2018koh, Chen:2019ykv}, in $e^+e^-$ annihilation~\cite{Jiang:2012jt, Jiang:2013ej, Zheng:2015ixa, Niu:2023ojf, Li:2026abl}, and in photon- or lepton-induced reactions~\cite{Li:2007vy, Chen:2014hqa, Bi:2016vbt, Sun:2020mvl, Zhan:2023fhh, Yang:2024ysg}. Indirect channels include the decays of top quarks, Higgs bosons, $W$ bosons, and $Z$ bosons~\cite{Niu:2018ycb, Ma:2022cgt, Niu:2019xuq, Ma:2022ger, Zhang:2022jst, Luo:2022jxq, Tian:2023yhs}. The general conclusion of the LO analyses is that the $P$-wave contribution amounts to only a few percent of the $S$-wave one for the inputs adopted. However, at scales near the charm- and bottom-quark masses the strong coupling is not small, so that higher-order QCD corrections are sizable.

The observation of the doubly charmed baryon $\Xi_{cc}^{++}$ by LHCb~\cite{LHCb:2017iph}, followed by the measurements of its lifetime and mass~\cite{LHCb:2018zpl, LHCb:2019qed}, established the doubly charmed sector. LHCb subsequently reported the first observation of the isospin partner $\Xi_{cc}^{+}$ using Run-3 data~\cite{LHCb:2026xicc}. However, LHCb searches for the bottom-charmed baryons $\Xi_{bc}$ and $\Omega_{bc}$ have so far found no established signal~\cite{LHCb:2020iko, LHCb:2021xba, LHCb:2022vqr}. The spectroscopy needed to interpret such searches has been studied with potential models, heavy-quark symmetry, lattice QCD, and QCD sum rules~\cite{Kiselev:2001fw, Gershtein:2000nx, Ebert:2002ig, Karliner:2014gca}. Lattice calculations resolve both the ground-state and excited doubly charmed states~\cite{Padmanath:2015jea}, while an explicit NLO QCD-sum-rule calculation shows that radiative corrections can materially reduce the mass-scheme and scale dependence~\cite{Wang:2017mqp}.

General NLO ingredients for heavy quarkonium production and decay were developed in Ref.~\cite{Petrelli:1997ge}; large corrections were subsequently found in hadroproduction~\cite{Campbell:2007ws, Butenschoen:2010rq, Ma:2010yw, Gong:2013qka}. For the $B_c^{(*)}$ system, NLO calculations have been carried out for production in top-quark decays~\cite{Sun:2010tx}, $Z^0$-boson decays~\cite{Qiao:2011zj, Jiang:2015pua, Ke:2026BcNLO}, and $e^+e^-$ annihilation~\cite{Zheng:2017xgj, Berezhnoy:2016hld}. Two independent studies of inclusive $B_c^{(*)}$ production in $W^{+}$ decays also found sizable QCD corrections~\cite{Chen:2019hck, Zheng:2019gnb}. The LHC provides a favorable experimental platform for studying the decay of the $W$ boson: about $3.07\times10^{10}$ $W$ bosons can be produced there per operation year~\cite{Qiao:2011yk}.

For the production of doubly heavy baryons through $W^{+}$ decays, $W^{+}\to\Xi_{Qc}+\bar{Q}+\bar{s}+X$ with $Q=b$ or $c$, the corresponding NLO short-distance coefficients are technically demanding to evaluate. The $\mathrm{SU}(3)_{C}$ color decomposition $\mathbf{3}\otimes\mathbf{3}=\bar{\mathbf{3}}\oplus\mathbf{6}$ splits the $Qc$ diquark into an antitriplet and a sextet channel, whose net color charges differ and therefore lead to different color correlations with the two accompanying antiquarks and, consequently, to different soft-gluon radiation patterns. The sextet channel moreover carries a distinct nonperturbative hadronization matrix element, which is poorly constrained and factorizes from the perturbative coefficient, so that it must be treated separately from the NLO QCD correction. The LO width of $W^{+}\to\Xi_{Qc}+\bar{Q}+\bar{s}+X$ is of $\mathcal{O}(\alpha\alpha_{s}^{2})$ and the NLO QCD correction of $\mathcal{O}(\alpha\alpha_{s}^{3})$. As will be shown below, the NLO prediction retains a strong dependence on the renormalization scale $\mu_{R}$ when the conventional scale-setting approach is used. Using the principle of maximum conformality (PMC) scale-setting method~\cite{Brodsky:2011ta, Mojaza:2012mf, Brodsky:2013vpa, Shen:2017pdu}, the non-conformal $\beta$-terms that drive the renormalization-group running of $\alpha_{s}$ are resummed into the argument of the coupling; the resulting PMC scale then reflects the physical virtuality of the gluons exchanged in the process. This yields a conformal perturbative series and, in turn, predictions free of the conventional scheme and scale ambiguities~\cite{Wu:2013ei, Wu:2014iba, Wu:2019mky, DiGiustino:2023jiq}.

The paper is organized as follows. In Sec.~\ref{sec:formalism}, we describe the calculation technology for the LO and NLO QCD corrections, with the derivation of the soft integrals collected in Appendix~\ref{app:soft-derivation}. In Sec.~\ref{sec:numerical}, we present the numerical results, including the total decay widths, the differential distributions, and the theoretical uncertainties induced by the heavy-quark masses and the renormalization scale; a PMC treatment of the NLO series is also presented there. Sec.~\ref{sec:summary} is devoted to a summary.

\section{Calculation technology}

\label{sec:formalism}

\subsection{Diquark factorization at LO}

Within the framework of NRQCD, the inclusive differential decay width for the process $W^{+}(p_{0})\to\Xi_{Qc}(p_{1})+\bar{Q}(p_{2})+\bar{s}(p_{3})+X$ can be written as
\begin{equation}
\mathrm{d}\Gamma=\sum_{n}\mathrm{d}\hat{\Gamma}_{W^{+}\to\langle Qc\rangle[n]+\bar{Q}+\bar{s}+X}\,\langle\mathcal{O}^{\Xi_{Qc}}[n]\rangle,
\label{factorization}
\end{equation}
where $Q=c$ and $b$ correspond to the doubly heavy baryons $\Xi_{cc}$ and $\Xi_{bc}$, respectively. $\langle Qc\rangle[n]$ labels the intermediate diquark state with the spin and color quantum numbers $[n]$. The long-distance matrix element $\langle\mathcal{O}^{\Xi_{Qc}}[n]\rangle$ describes the nonperturbative transition probability of a $\langle Qc\rangle[n]$ state into the observed baryon $\Xi_{Qc}$. For an attractive color-antitriplet $S$-wave state, it is commonly approximated by $|R(0)|^{2}/(4\pi)$. No equally direct potential-model interpretation exists for the repulsive sextet states. Following common phenomenological practice, we adopt $h_{\mathbf{6}}=h_{\bar{\mathbf{3}}}$ in the numerical calculations below~\cite{Ma:2003zk}.

\begin{figure}[H]
\centering
\includegraphics[width=0.26\textwidth]{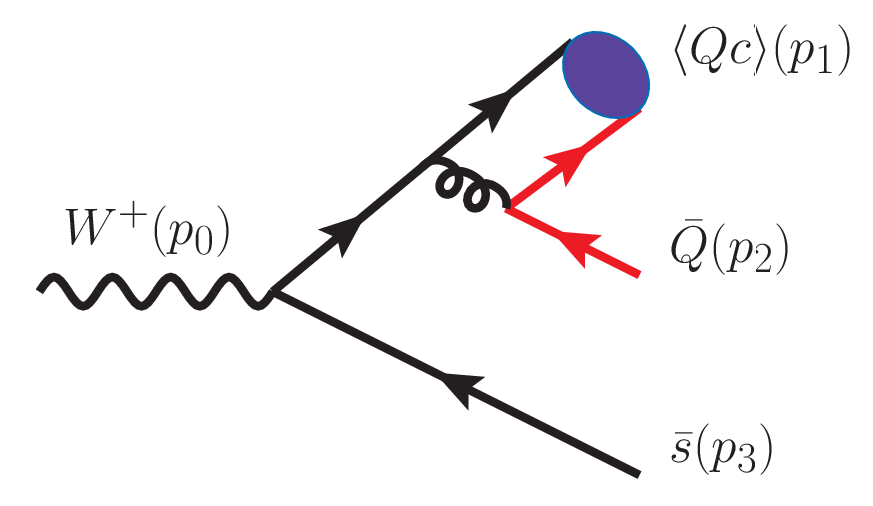}
\includegraphics[width=0.26\textwidth]{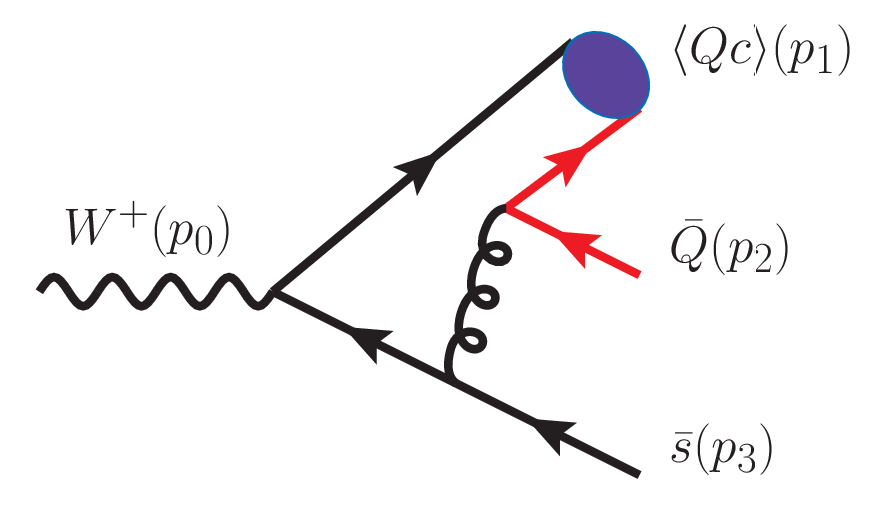}\\[3pt]
\includegraphics[width=0.25\textwidth]{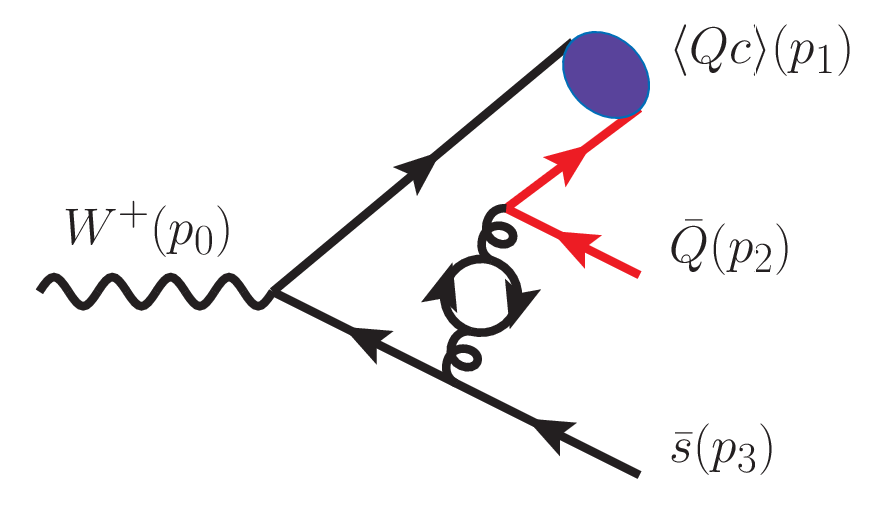}\hfill
\includegraphics[width=0.25\textwidth]{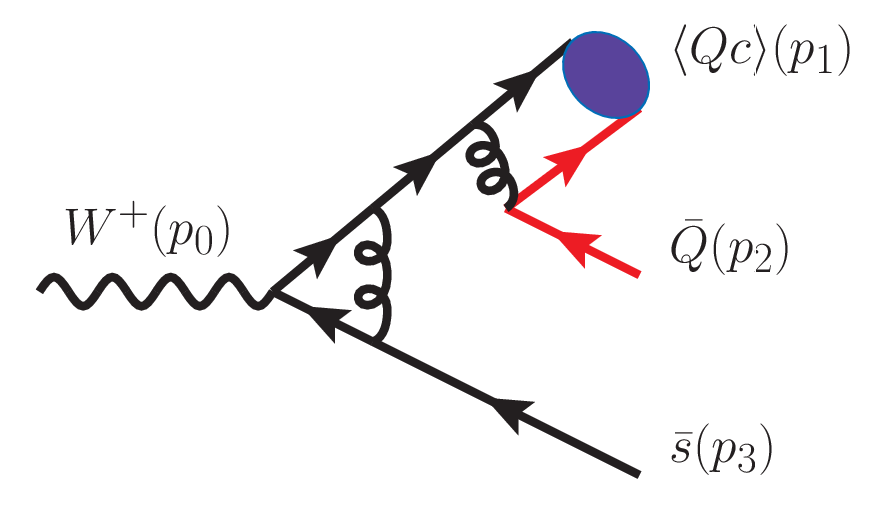}\hfill
\includegraphics[width=0.25\textwidth]{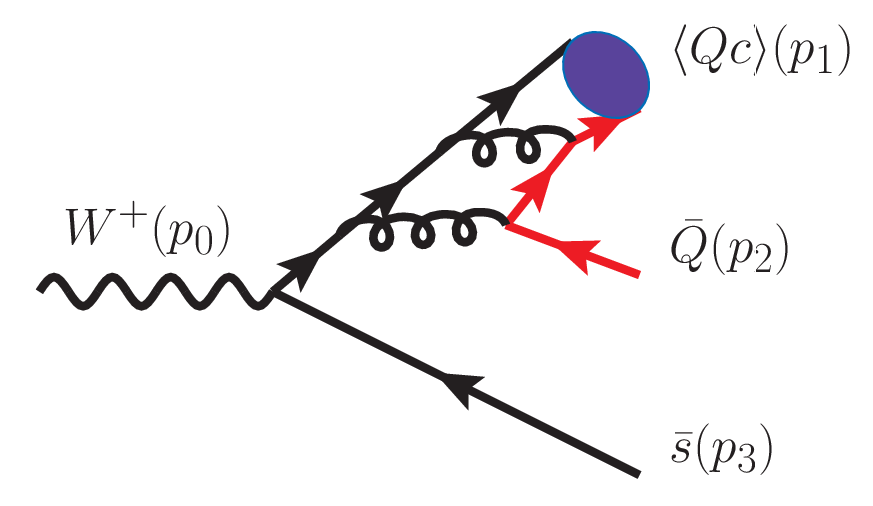}\hfill
\includegraphics[width=0.25\textwidth]{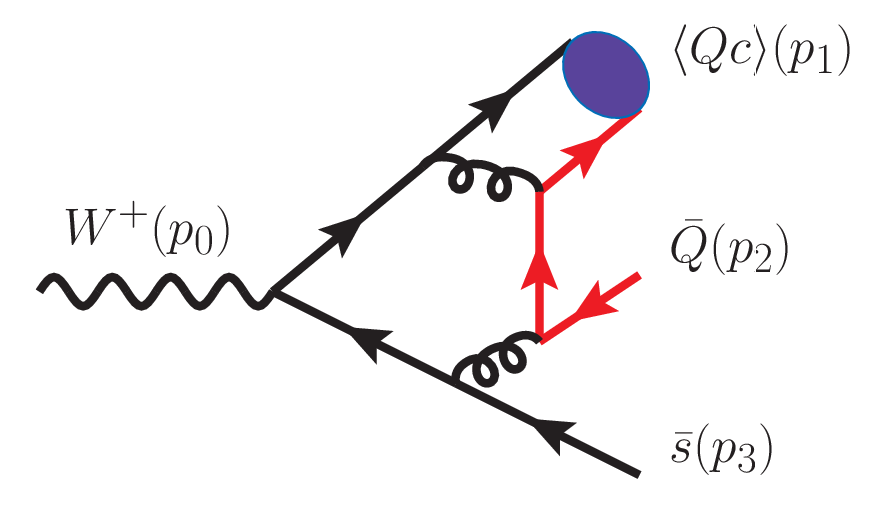}\\[3pt]
\includegraphics[width=0.25\textwidth]{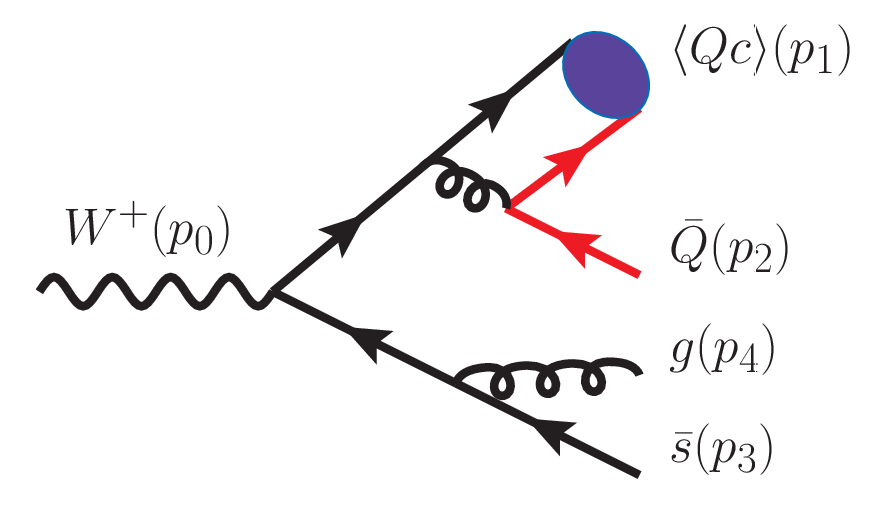}\hfill
\includegraphics[width=0.25\textwidth]{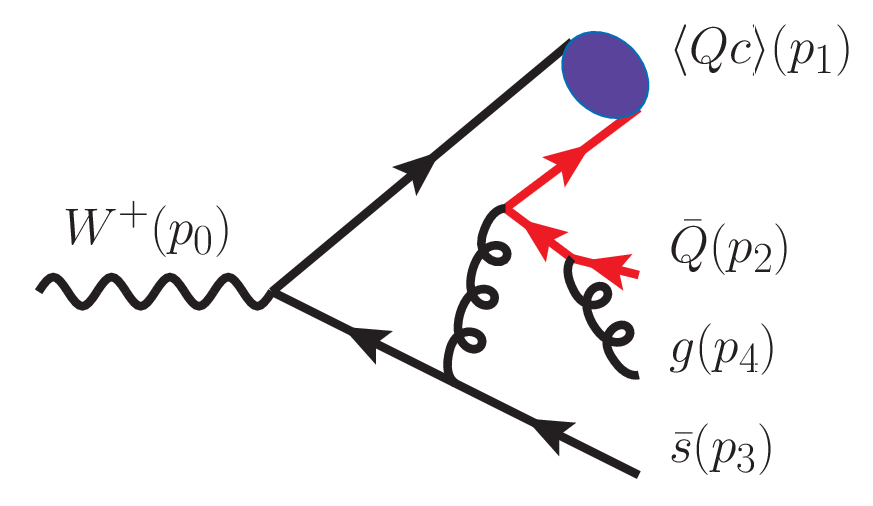}\hfill
\includegraphics[width=0.25\textwidth]{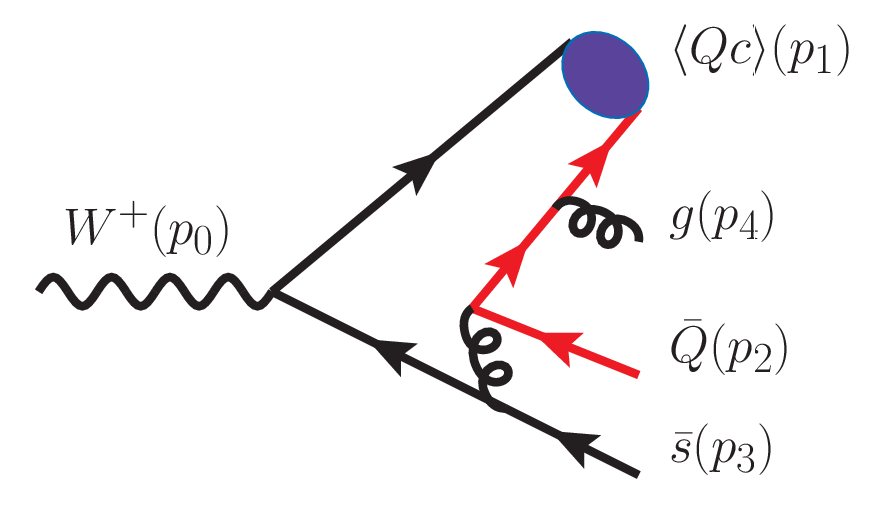}\hfill
\includegraphics[width=0.25\textwidth]{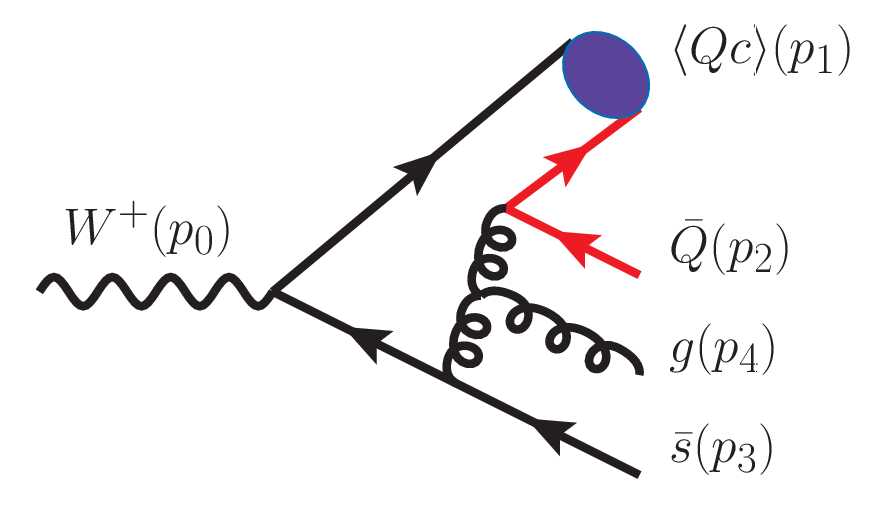}\\[3pt]
\caption{Typical Feynman diagrams for $W^{+}(p_{0})\to\langle Qc\rangle[n](p_{1})+\bar{Q}(p_{2})+\bar{s}(p_{3})$ with $Q=c$ or $b$. From top to bottom, the rows show the Born, virtual, and real-emission diagrams. The red lines mark the $Q\bar{Q}$ pair, whose fermion-flow direction is reversed by charge conjugation.}
\label{fig:born-diagrams}
\end{figure}

In the hard-scattering process, two typical Feynman diagrams contribute to the production of $\Xi_{Qc}$ through $W^{+}$ decays; they are shown in the first two panels of Fig.~\ref{fig:born-diagrams}, with the $+X$ part omitted. As a subtlety in the differential decay width, an extra factor of $2^{2}/2!$ must be included for $\Xi_{cc}$ because of the two identical charm quarks in the diquark system. The charge-conjugation matrix $C=-i\gamma^{2}\gamma^{0}$ has to be applied to reverse the $Q\bar{Q}$ fermion chain (the red line in Fig.~\ref{fig:born-diagrams}), which yields the diquark amplitude up to an extra factor $(-1)^{n+1}$, with $n$ the number of vector vertices in the red fermion chain~\cite{Zheng:2015ixa, Chen:2014hqa}. The momenta of the constituent quarks inside the diquark read $p_{11}=\frac{m_{Q}}{m_{Q}+m_{c}}\,p_{1}-q$ and $p_{12}=\frac{m_{c}}{m_{Q}+m_{c}}\,p_{1}+q$, where $q$ is the relative momentum of the two constituent quarks and can be set to zero ($q\to0$) for $S$-wave states. The spin projectors for the spin-singlet and spin-triplet states are
\begin{align}
\Pi[{}^{1}S_{0}](p_{1}) &= -\frac{\sqrt{M}}{4m_{Q} m_{c}}\left(\not{p}_{12}-m_{c}\right)\gamma_{5}\left(\not{p}_{11}+m_{Q}\right), \\
\Pi[{}^{3}S_{1}](p_{1}) &= -\frac{\sqrt{M}}{4m_{Q} m_{c}}\left(\not{p}_{12}-m_{c}\right)\not{\varepsilon}(p_{1})\left(\not{p}_{11}+m_{Q}\right),
\end{align}
where $\varepsilon(p_{1})$ is the polarization vector of the spin-triplet diquark. We take $M=m_{Q}+m_{c}$ to leading order in the relative velocity $v$. For the color factor, we introduce the color tensors $G_{mjk}$ normalized by $\mathcal{N}=\sqrt{1/2}$, i.e., $G_{mjk}=\mathcal{N}\epsilon_{mjk}$ for the color-antitriplet channel and $G_{mjk}=\mathcal{N}f_{mjk}$ for the color-sextet channel. The antisymmetric function $\epsilon_{mjk}$ and the symmetric function $f_{mjk}$ satisfy
\begin{align}
\epsilon_{mjk}\epsilon_{m'j'k} &= \delta_{mm'}\delta_{jj'}-\delta_{mj'}\delta_{jm'}, \\
f_{mjk}f_{m'j'k} &= \delta_{mm'}\delta_{jj'}+\delta_{mj'}\delta_{jm'}.
\end{align}
Contracting $m$ and $j$, these reduce to $\sum_{m,j}\epsilon_{mjk}\epsilon_{mjk'}=2\delta_{kk'}$ and $\sum_{m,j}f_{mjk}f_{mjk'}=2\delta_{kk'}$, so that the color tensors are orthonormal,
\begin{equation}
\sum_{m,j}G_{mjk}G_{mjk'}=\delta_{kk'}.
\end{equation}

The LO Feynman amplitude $\mathcal{M}_{\mathrm{LO}}$ can be read off directly from the Feynman rules, following the diagrams in the first row of Fig.~\ref{fig:born-diagrams}. We generate Feynman diagrams and the partonic amplitudes with \textsc{FeynArts}~\cite{Hahn:2000kx} and adopt \textsc{FeynCalc}~\cite{Mertig:1990an} to carry out the Dirac and color traces. The corresponding LO differential decay width $\mathrm{d}\hat{\Gamma}_{\mathrm{LO}}$ and the three-body phase space $\mathrm{d}\Phi_{3}$ are expressed as
\begin{align}
\mathrm{d}\hat{\Gamma}_{\mathrm{LO}} &= \frac{1}{3}\frac{1}{2m_{W}}\sum\left|\mathcal{M}_{\mathrm{LO}}\right|^{2}\mathrm{d}\Phi_{3}, \\
\mathrm{d}\Phi_{3} &= (2\pi)^{D}\delta^{D}\Big(p_{0}-\sum_{f=1}^{3}p_{f}\Big)\prod_{f=1}^{3}\frac{\mathrm{d}^{D-1}\mathbf{p}_{f}}{(2\pi)^{D-1}2E_{f}},
\end{align}
where $\sum$ denotes the sum over the color and spin states of the initial and final particles, the factor $1/3$ accounts for the spin average of the initial $W^{+}$ boson, and $D$ is the number of spacetime dimensions.

\subsection{NLO QCD corrections}

The NLO QCD corrections consist of both virtual and real contributions. The virtual contribution to the differential decay width reads
\begin{equation}
\mathrm{d}\hat{\Gamma}_{\mathrm{Virtual}}=\frac{1}{3}\frac{1}{2m_{W}}\sum 2\,\mathrm{Re}\left(\mathcal{M}_{\mathrm{LO}}^{*}\mathcal{M}_{\mathrm{Virtual}}\right)\mathrm{d}\Phi_{3},
\end{equation}
where $\mathcal{M}_{\mathrm{Virtual}}$ is the amplitude of the virtual corrections. Representative virtual QCD topologies for $W^{+}\to\langle Qc\rangle[n]+\bar{Q}+\bar{s}$ are shown in the second row of Fig.~\ref{fig:born-diagrams}. For each projected one-loop amplitude, we form its interference with $\mathcal{M}_{\mathrm{LO}}$ before reducing the loop integrals. 

The virtual corrections contain both ultraviolet (UV) and infrared (IR) divergences. The UV ones are removed by renormalization. Using the counterterm approach, the renormalized quark mass $m$, quark field $\Psi_{r}$, gluon field $A_{r}^{\mu}$, and strong coupling constant $g_{s}$ are introduced as
\begin{align}
m_{0} &= Z_{m}m, & \Psi_{0} &= \sqrt{Z_{2}}\,\Psi_{r}, \\
A_{0}^{\mu} &= \sqrt{Z_{3}}\,A_{r}^{\mu}, & g_{s0} &= Z_{g}g_{s},
\end{align}
where $Z_{i}=1+\delta Z_{i}$ $(i=m,2,3,g)$ are the renormalization constants, with
\begin{align*}
\delta Z^{\mathrm{OS}}_{m,Q} &= -3C_{F}\frac{\alpha_{s}}{4\pi}\left[\frac{1}{\varepsilon_{\mathrm{UV}}}-\gamma_{E}+\ln\frac{4\pi\mu_{R}^{2}}{m_{Q}^{2}}+\frac{4}{3}\right],\\
\delta Z^{\mathrm{OS}}_{2,Q} &= -C_{F}\frac{\alpha_{s}}{4\pi}\left[\frac{1}{\varepsilon_{\mathrm{UV}}}+\frac{2}{\varepsilon_{\mathrm{IR}}}-3\gamma_{E}+3\ln\frac{4\pi\mu_{R}^{2}}{m_{Q}^{2}}+4\right],\\
\delta Z^{\mathrm{OS}}_{2,q} &= -C_{F}\frac{\alpha_{s}}{4\pi}\left[\frac{1}{\varepsilon_{\mathrm{UV}}}-\frac{1}{\varepsilon_{\mathrm{IR}}}\right],\\
\delta Z^{\mathrm{OS}}_{3} &= \frac{\alpha_{s}}{4\pi}\left\{(\beta_{0}^{\prime}-2C_{A})\left(\frac{1}{\varepsilon_{\mathrm{UV}}}-\frac{1}{\varepsilon_{\mathrm{IR}}}\right)\right.\nonumber\\
&\quad\left.-\frac{4}{3}T_{F}\left(\frac{1}{\varepsilon_{\mathrm{UV}}}-\gamma_{E}+\ln\frac{4\pi\mu_{R}^{2}}{m_{c}^{2}}\right)-\frac{4}{3}T_{F}\left(\frac{1}{\varepsilon_{\mathrm{UV}}}-\gamma_{E}+\ln\frac{4\pi\mu_{R}^{2}}{m_{b}^{2}}\right)\right\},\\
\delta Z^{\overline{\mathrm{MS}}}_{g} &= -\frac{\beta_{0}}{2}\frac{\alpha_{s}}{4\pi}\left[\frac{1}{\varepsilon_{\mathrm{UV}}}-\gamma_{E}+\ln(4\pi)\right],
\end{align*}
where $\mathrm{OS}$ and $\overline{\mathrm{MS}}$ in the superscripts indicate the renormalization schemes. The quark masses, the heavy- and light-quark fields, and the gluon field are renormalized in the on-shell scheme, while the strong coupling constant is renormalized in the $\overline{\mathrm{MS}}$ scheme. $Q=c,b$ in the subscripts denotes the heavy quark and $q=s$ the light quark. $\mu_{R}$ is the renormalization scale. In QCD, $C_{A}=3$, $C_{F}=4/3$, and $T_{F}=1/2$. $\beta_{0}=11-2n_{f}/3$ is the one-loop coefficient of the $\beta$ function for $n_{f}$ active flavors. In $\delta Z^{\mathrm{OS}}_{3}$, $\beta_{0}^{\prime}=11C_{A}/3-4T_{F}n_{l}/3$, where $n_{l}=3$ counts the massless $u$, $d$, and $s$ flavors; the heavy quarks' contributions are displayed separately.

The IR-safe result is obtained by combining the virtual contribution with the real corrections, the latter treated with the two-cutoff phase-space slicing method~\cite{Harris:2001sx}. Conventional dimensional regularization, $D=4-2\varepsilon$, is employed to isolate these divergences. To simplify the calculation, we use the naive anticommuting prescription for $\gamma_{5}$ in $D$ dimensions,
\begin{equation}
\{\gamma_{5},\gamma^{\mu}\}=0,\qquad \gamma_{5}^{2}=1.
\label{eq:gamma5}
\end{equation}
The $\gamma_{5}$ factors arise from the weak vertex and from the spin-singlet diquark projector. In the \textsc{FeynCalc} evaluation, the $\gamma_{5}$ matrices along each Dirac chain are first reduced in pairs, since $\gamma_{5}^{2}=1$ and $\gamma_{5}$ anticommutes with $\gamma^{\mu}$. An even number of $\gamma_{5}$ factors is therefore removed entirely before the remaining trace is evaluated, whereas for an odd number the pairs are removed in the same way, leaving only a single $\gamma_{5}$ in place; the trace is then evaluated with that one factor kept, using $\mathrm{tr}[\gamma_{5}\gamma^{\mu}\gamma^{\nu}\gamma^{\rho}\gamma^{\sigma}]=-4i\epsilon^{\mu\nu\rho\sigma}$ (and vanishing for fewer than four $\gamma$ matrices).

After the Dirac traces, the loop numerators are expressed in terms of scalar products. Then we use the \texttt{Apart} package~\cite{Feng:2012iq} to perform partial-fraction decomposition and the \textsc{FIRE} package~\cite{Smirnov:2008iw} to reduce the resulting loop integrals, using integration-by-parts (IBP) identities, to the master integrals, namely the $A_0$, $B_0$, $C_0$, and $D_0$ functions. The master integrals can be evaluated numerically using \textsc{LoopTools}~\cite{Hahn:1998yk}.

The decay width for real gluon emission, $W^{+}(p_{0})\to\langle Qc\rangle[n](p_{1})+\bar{Q}(p_{2})+\bar{s}(p_{3})+g(p_{4})$, can be expressed in terms of the four-body phase space $\mathrm{d}\Phi_{4}$,
\begin{align}
\mathrm{d}\hat{\Gamma}_{\mathrm{Real}} &= \frac{1}{3}\frac{1}{2m_{W}}\sum\left|\mathcal{M}_{\mathrm{Real}}\right|^{2}\mathrm{d}\Phi_{4}, \\
\mathrm{d}\Phi_{4} &= (2\pi)^{D}\delta^{D}\Big(p_{0}-\sum_{f=1}^{4}p_{f}\Big)\prod_{f=1}^{4}\frac{\mathrm{d}^{D-1}\mathbf{p}_{f}}{(2\pi)^{D-1}2E_{f}}.
\end{align}
The representative real-emission topologies for $W^{+}(p_{0})\to\langle Qc\rangle[n](p_{1})+\bar{Q}(p_{2})+\bar{s}(p_{3})+g(p_{4})$ are shown in the third row of Fig.~\ref{fig:born-diagrams}. In the real corrections, two small cutoffs, $\delta_{s}$ and $\delta_{c}$ ($\delta_{c}\ll\delta_{s}$), are introduced to divide the real phase space into three regions: soft (S), hard-collinear (HC), and hard-noncollinear ($\mathrm{H}\bar{\mathrm{C}}$). The cutoff $\delta_{s}$ separates the soft ($E_{4}\le m_{W}\delta_{s}/2$) from the hard ($E_{4}>m_{W}\delta_{s}/2$) region, where $E_{4}$ is the gluon energy in the rest frame of the initial $W^{+}$ boson. Within the hard region, $\delta_{c}$ further separates the collinear configuration ($(p_{3}+p_{4})^{2}\le\delta_{c}m_{W}^{2}$) from the hard-noncollinear one ($\mathrm{H}\bar{\mathrm{C}}$, with $(p_{3}+p_{4})^{2}>\delta_{c}m_{W}^{2}$). The $\mathrm{H}\bar{\mathrm{C}}$ region is finite and can be integrated numerically in four dimensions.

At leading order in the relative velocity, the two heavy constituents carry momenta proportional to $p_{1}$, and their eikonal currents combine into a single current with color charge $\mathbf{T}_{1}=\mathbf{T}_{Q}+\mathbf{T}_{c}$. For a fixed diquark color representation $R$ ($=\bar{\mathbf{3}}$ or $\mathbf{6}$) and soft-gluon momentum $k=p_{4}$, the squared amplitude is
\begin{align*}
\sum\left|\mathcal{M}_{\mathrm{Real}}^{\mathrm{soft}}\right|^{2}
&= -g_{s}^{2}\mu_{R}^{2\varepsilon}\sum_{i,j=1}^{3}\mathcal{C}_{ij}^{(R)}
\frac{p_{i}\cdot p_{j}}{(p_{i}\cdot k)(p_{j}\cdot k)}
\sum\left|\mathcal{M}_{\mathrm{LO}}^{R}\right|^{2}.
\end{align*}
Here $\mathcal{C}_{ij}^{(R)}$ is the normalized color correlation defined in Appendix~\ref{app:soft-derivation}. Then the soft (S) contribution is
\begin{equation}
\mathrm{d}\hat{\Gamma}_{S}^{R}=\mathrm{d}\hat{\Gamma}_{\mathrm{LO}}^{R}
\frac{C_{F}\alpha_{s}}{2\pi}\frac{\Gamma(1-\varepsilon)}{\Gamma(1-2\varepsilon)}
\left(\frac{4\pi\mu_{R}^{2}}{m_{W}^{2}}\right)^{\varepsilon}S_{R},
\label{eq:soft-width}
\end{equation}
where the integrated soft factors are
\begin{equation}
\begin{aligned}
S_{\bar{\mathbf{3}}} &= -I_{11}-I_{22}+I_{12}+I_{13}+I_{23},\\
S_{\mathbf{6}} &= -\frac{5}{2}I_{11}-I_{22}+\frac{5}{2}I_{12}+\frac{5}{2}I_{13}-\frac{1}{2}I_{23}.
\end{aligned}
\label{eq:soft-unified}
\end{equation}
The color algebra and the complete expressions for the dimensionless integrals $I_{ij}$ are given in Appendix~\ref{app:soft-derivation}. The $\Gamma$-function prefactors and the scale factor in Eq.~\eqref{eq:soft-width} are kept unexpanded in $\varepsilon$.

The hard-collinear (HC) pole arises solely from the emission of a gluon collinear to the massless $\bar{s}$ antiquark; the corresponding formula is the same as that in Ref.~\cite{Zheng:2019gnb}, which implies
\begin{equation}
\mathrm{d}\hat{\Gamma}_{\mathrm{HC}}^{R}=\mathrm{d}\hat{\Gamma}_{\mathrm{LO}}^{R}(D)
\frac{C_{F}\alpha_{s}}{2\pi}\frac{\Gamma(1-\varepsilon)}{\Gamma(1-2\varepsilon)}
\left(\frac{4\pi\mu_{R}^{2}}{m_{W}^{2}}\right)^{\varepsilon}
\left[\frac{A_{1}}{\varepsilon}+A_{0}\right],
\label{eq:hc-width}
\end{equation}
where
\begin{align}
A_{1} &= \frac{3}{2}+2\ln\delta_{s}^{\prime},\nonumber\\
A_{0} &= \frac{7}{2}-\frac{\pi^{2}}{3}-\ln^{2}\delta_{s}^{\prime}
-\ln\delta_{c}\left(\frac{3}{2}+2\ln\delta_{s}^{\prime}\right),
\label{eq:hc-coefficients}
\end{align}
with
\begin{equation}
\delta_{s}^{\prime}=\frac{m_{W}^{2}\delta_{s}}{m_{W}^{2}-(p_{1}+p_{2})^{2}}.
\label{eq:deltasp}
\end{equation}
These HC coefficients are identical for both diquark color representations, since the dependence on the color state enters only through the Born factor.

The soft and hard-collinear contributions separately contain IR poles and logarithms of $\delta_{s}$ and $\delta_{c}$; these cancel only after the virtual, soft, hard-collinear, and hard-noncollinear pieces are combined, so that the sum is independent of $\delta_{s}$ and $\delta_{c}$. For definiteness, we choose $\delta_{s}=1\times10^{-6}$ and $\delta_{c}=2\times10^{-8}$.

The NLO correction $\Delta\Gamma_{\mathrm{NLO}}=\Gamma_{\mathrm{Real}}+\Gamma_{\mathrm{Virtual}}$ is of $\mathcal{O}(\alpha\alpha_{s}^{3})$. $\Gamma_{\mathrm{tot}}$ denotes the decay width including the NLO correction,
\begin{equation}
\Gamma_{\mathrm{tot}}=\Gamma_{\mathrm{LO}}+\Delta\Gamma_{\mathrm{NLO}}.
\end{equation}

\section{Numerical results}
\label{sec:numerical}

In the numerical calculations, the relevant input parameters are taken as follows~\cite{ParticleDataGroup:2026}:
\begin{align}
&m_{Z}=91.1876~\mathrm{GeV},\quad m_{W}=80.4~\mathrm{GeV}, \nonumber\\
&\Gamma_{W}=2.085~\mathrm{GeV},\quad G_{F}=1.16638\times10^{-5}~\mathrm{GeV}^{-2}, \nonumber\\
&m_{c}=1.8~\mathrm{GeV},\quad m_{b}=5.1~\mathrm{GeV},\quad |V_{cs}|=1.
\end{align}
Taking $\alpha_{s}(m_{Z})=0.118$ as the reference value, we obtain $\Lambda_{\mathrm{QCD}}^{(n_{f}=4)}=0.330~\mathrm{GeV}$ and $\Lambda_{\mathrm{QCD}}^{(n_{f}=5)}=0.226~\mathrm{GeV}$. The running couplings $\alpha_{s}^{(n_{f}=4)}(2m_{c})=0.239$ and $\alpha_{s}^{(n_{f}=5)}(2m_{b})=0.177$ are obtained by solving the two-loop renormalization group equation. The default renormalization scale for the production of $\Xi_{cc}$ ($\Xi_{bc}$) is taken as $\mu_{0}=2m_{c}$ ($\mu_{0}=2m_{b}$). The long-distance matrix elements are taken as $h_{\mathbf{6}}=h_{\bar{\mathbf{3}}}=0.039~\mathrm{GeV}^{3}$ for the $cc$ system and $0.065~\mathrm{GeV}^{3}$ for the $bc$ system, where $h_{\bar{\mathbf{3}}}$ follows the potential-model estimate~\cite{Kiselev:2001fw, Gershtein:2000nx}, while $h_{\mathbf{6}}$ is set equal to it as assumed above. 

\subsection{Basic results}

At LO and NLO, the decay widths for the production of $\Xi_{cc}$ and $\Xi_{bc}$ through $W^{+}\to\Xi_{cc}+\bar{c}+\bar{s}+X$ and $W^{+}\to\Xi_{bc}+\bar{b}+\bar{s}+X$ are presented in Table~\ref{tab:decaywidth}, where the $K$ factor is defined as $K=\Gamma_{\mathrm{tot}}/\Gamma_{\mathrm{LO}}$. The NLO effects are sizable: the total $K$ factors are $2.27$ for $\Xi_{cc}$ and $2.95$ for $\Xi_{bc}$. For the same-spin $\langle bc\rangle$ channels, the LO squared color factors are $4/3$ for the color antitriplet $\bar{\mathbf{3}}$ and $2/3$ for the color sextet $\mathbf{6}$~\cite{Zhang:2022jst}. With the assumption $h_{\mathbf{6}}=h_{\bar{\mathbf{3}}}$, the LO decay width of the $\mathbf{6}$ channel is therefore approximately one half of that of the $\bar{\mathbf{3}}$ channel. At NLO, however, the color structure no longer reduces to an overall multiplicative factor. For example, the coefficients of $I_{12}$ and $I_{13}$ in the soft factor $S_{R}$ are $C_{R}/C_{F}=5/2$ for $\mathbf{6}$ and $1$ for $\bar{\mathbf{3}}$, as shown in Eq.~\eqref{eq:app-S-dipoles}. Numerically, the NLO corrections of the two color states for $\langle bc\rangle[{}^{1}S_{0}]$ are comparable, $1.03$ and $1.01~\mathrm{keV}$ for $\bar{\mathbf{3}}$ and $\mathbf{6}$, respectively. For $\langle bc\rangle[{}^{3}S_{1}]$, the NLO correction of the color sextet is relatively large, being $1.80~\mathrm{keV}$ compared with $1.15~\mathrm{keV}$ for the antitriplet. A smaller LO width together with a comparable or larger NLO correction therefore yields a larger sextet $K$ factor in both spin channels. A larger $K$ factor also appears in the $\langle cc\rangle[{}^{1}S_{0}]_{\mathbf{6}}$ channel. These results motivate an NNLO calculation to assess the perturbative stability and to quantify the missing higher-order contributions.

\begin{table}[H]
\caption{Decay widths (in keV) and $K$ factors for the production of $\Xi_{cc}$ and $\Xi_{bc}$ through $W^{+}$ decays. Here the sums run over the channels listed and are computed from the unrounded values, so they may differ slightly from the sums of the displayed entries.}
\label{tab:decaywidth}
\centering
\begin{tabular*}{\textwidth}{@{\extracolsep{\fill}}lcccc@{}}
\hline\hline
State & $\Gamma_{\mathrm{LO}}$ & $\Delta\Gamma_{\mathrm{NLO}}$ & $\Gamma_{\mathrm{tot}}$ & $K$ \\
\hline
$\langle cc\rangle[{}^{1}S_{0}]_{\mathbf{6}}$ & $26.50$ & $51.83$ & $78.34$ & $2.96$ \\
$\langle cc\rangle[{}^{3}S_{1}]_{\bar{\mathbf{3}}}$ & $54.85$ & $51.58$ & $106.43$ & $1.94$ \\
$\langle cc\rangle$-sum & $81.36$ & $103.42$ & $184.77$ & $2.27$ \\
\hline
$\langle bc\rangle[{}^{1}S_{0}]_{\bar{\mathbf{3}}}$ & $0.91$ & $1.03$ & $1.94$ & $2.13$ \\
$\langle bc\rangle[{}^{1}S_{0}]_{\mathbf{6}}$ & $0.46$ & $1.01$ & $1.46$ & $3.21$ \\
$\langle bc\rangle[{}^{3}S_{1}]_{\bar{\mathbf{3}}}$ & $0.80$ & $1.15$ & $1.95$ & $2.45$ \\
$\langle bc\rangle[{}^{3}S_{1}]_{\mathbf{6}}$ & $0.40$ & $1.80$ & $2.20$ & $5.53$ \\
$\langle bc\rangle$-sum & $2.56$ & $5.00$ & $7.56$ & $2.95$ \\
\hline\hline
\end{tabular*}
\end{table}

With the definition $\mathcal{B}=\Gamma_{\mathrm{tot}}/\Gamma_{W}$, the inclusive branching ratios for $\Xi_{cc}$ and $\Xi_{bc}$ production through $W^{+}$ decays at NLO are
\begin{align}
\mathcal{B}(W^{+}\to\Xi_{cc}+\bar{c}+\bar{s}+X) &\approx 8.86\times10^{-5}, \\
\mathcal{B}(W^{+}\to\Xi_{bc}+\bar{b}+\bar{s}+X) &\approx 3.63\times10^{-6}.
\end{align}
Using the benchmark $N_{W^{+}}=3.07\times10^{10}$ $W^{+}$ bosons produced per operation year at the LHC~\cite{Qiao:2011yk, Zhang:2022jst}, we estimate that $2.72\times10^{6}$ $\Xi_{cc}$ baryons and $1.11\times10^{5}$ $\Xi_{bc}$ baryons can be produced per operation year.

\begin{figure}[H]
\centering
\includegraphics[width=0.49\textwidth]{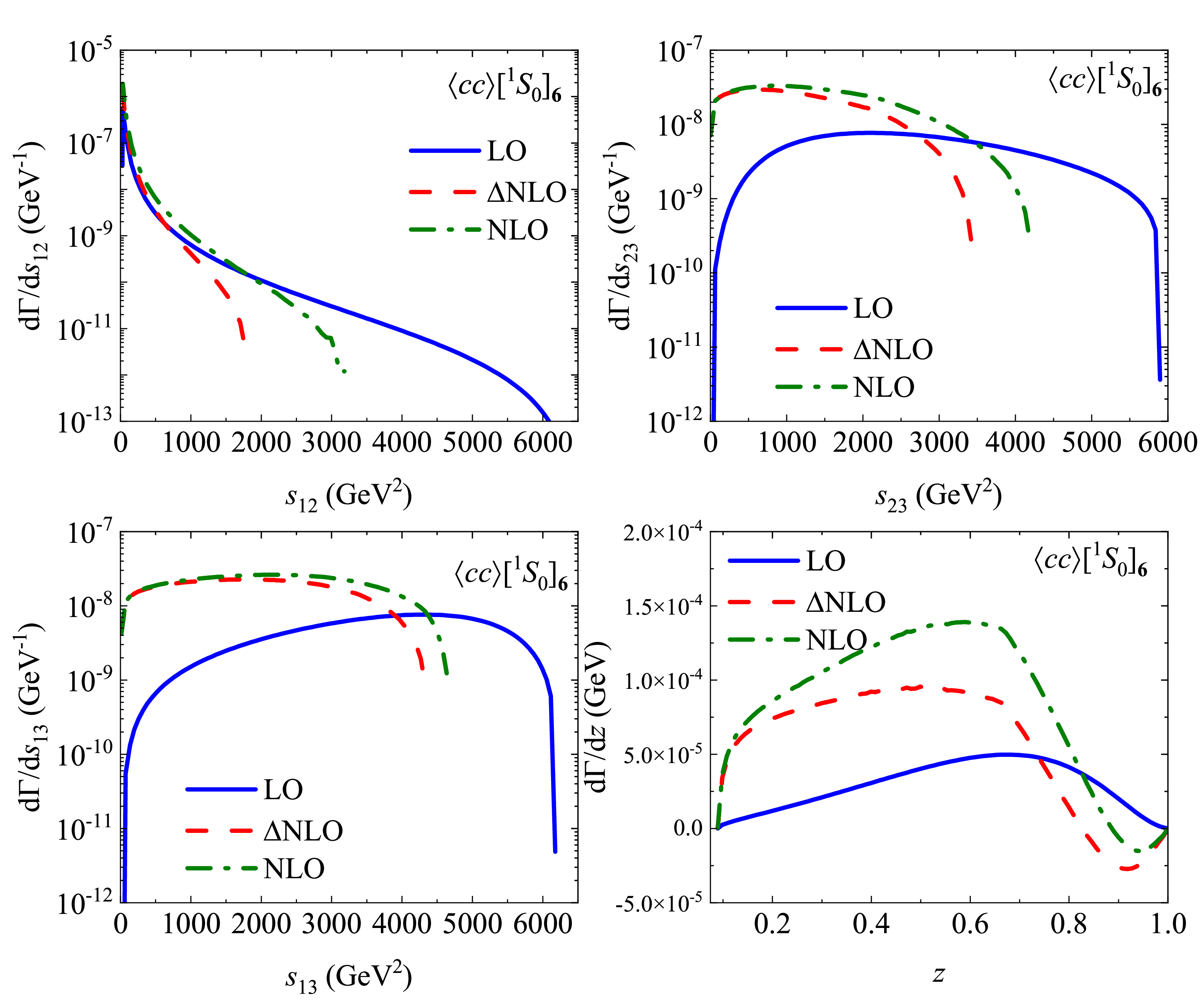}
\includegraphics[width=0.49\textwidth]{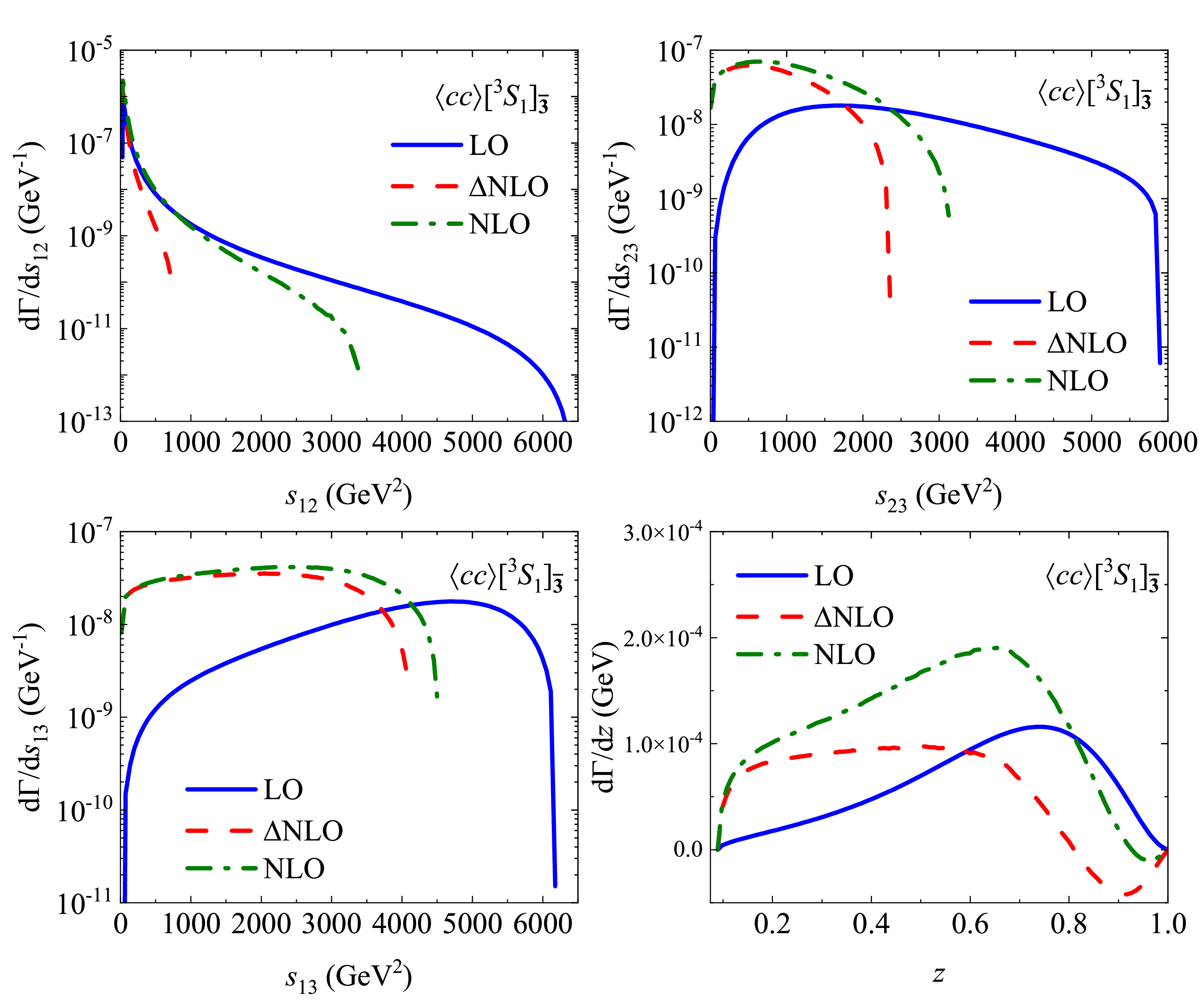}
\caption{Differential decay widths for $\langle cc\rangle[{}^{1}S_{0}]_{\mathbf{6}}$ (left) and $\langle cc\rangle[{}^{3}S_{1}]_{\bar{\mathbf{3}}}$ (right).}
\label{fig:Xicc_distributions}
\end{figure}

\begin{figure}[H]
\centering
\includegraphics[width=0.49\textwidth]{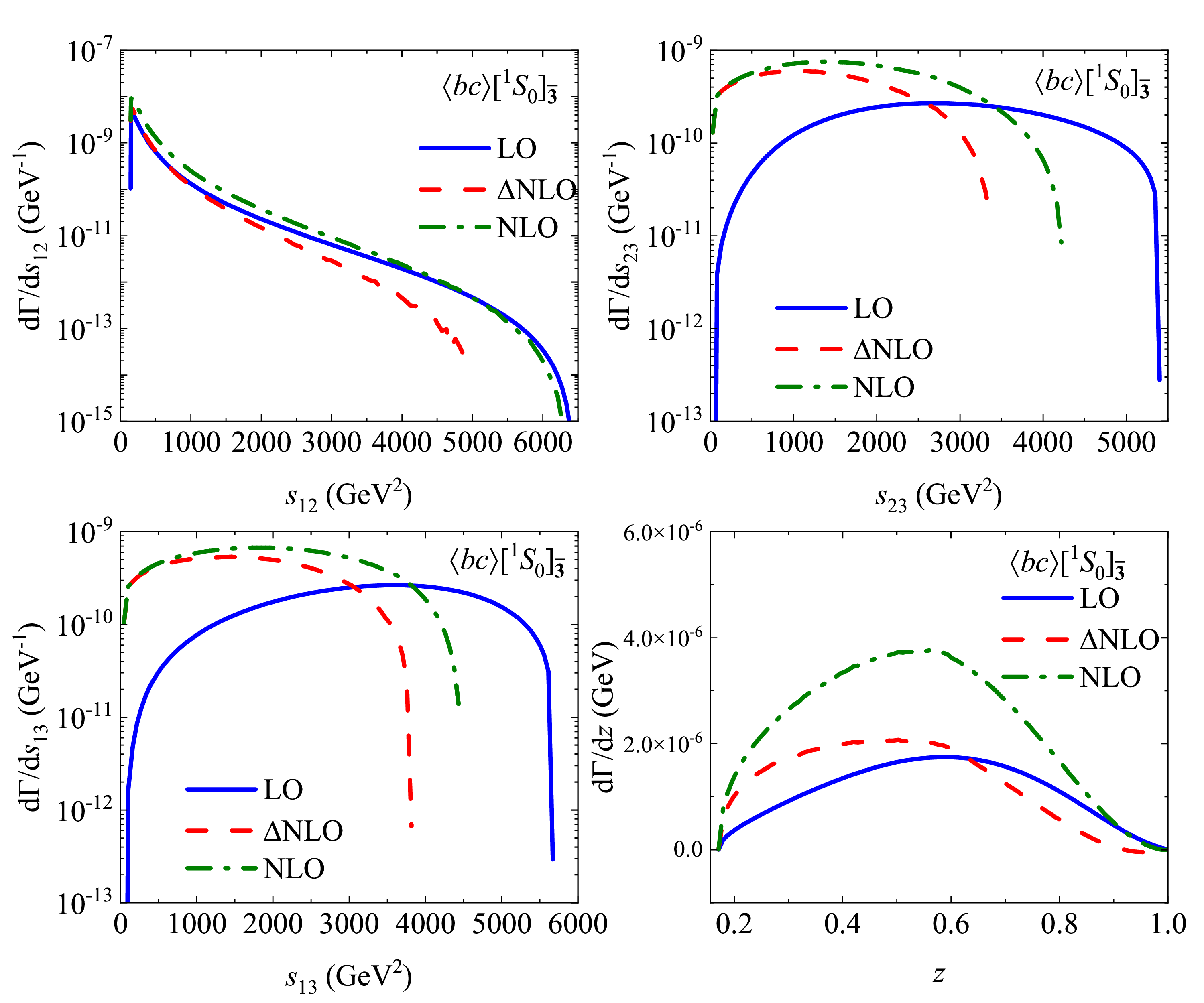}
\includegraphics[width=0.49\textwidth]{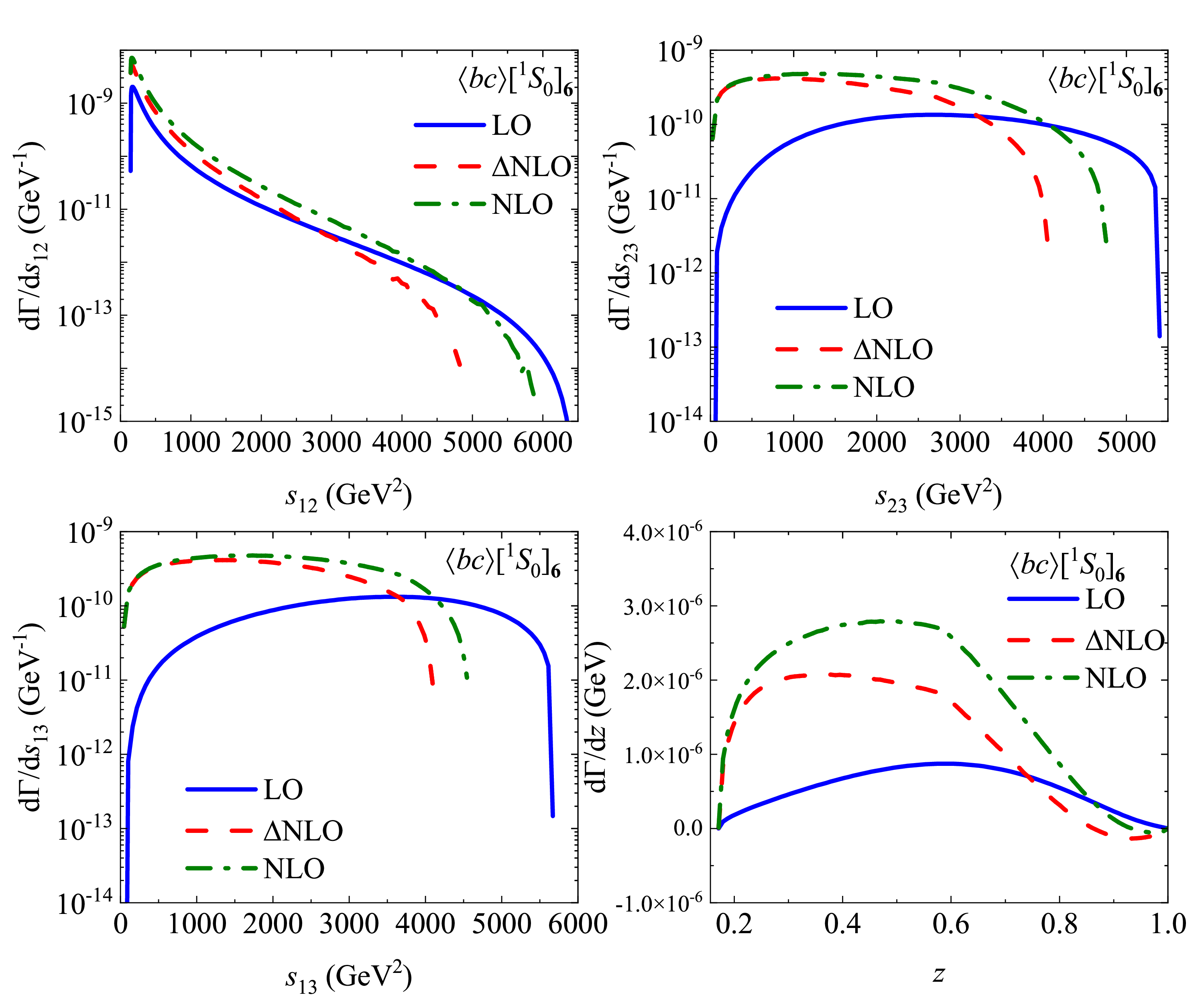}
\caption{Differential decay widths for $\langle bc\rangle[{}^{1}S_{0}]_{\bar{\mathbf{3}}}$ (left) and $\langle bc\rangle[{}^{1}S_{0}]_{\mathbf{6}}$ (right).}
\label{fig:Xibc_1s0_distributions}
\end{figure}

\begin{figure}[H]
\centering
\includegraphics[width=0.49\textwidth]{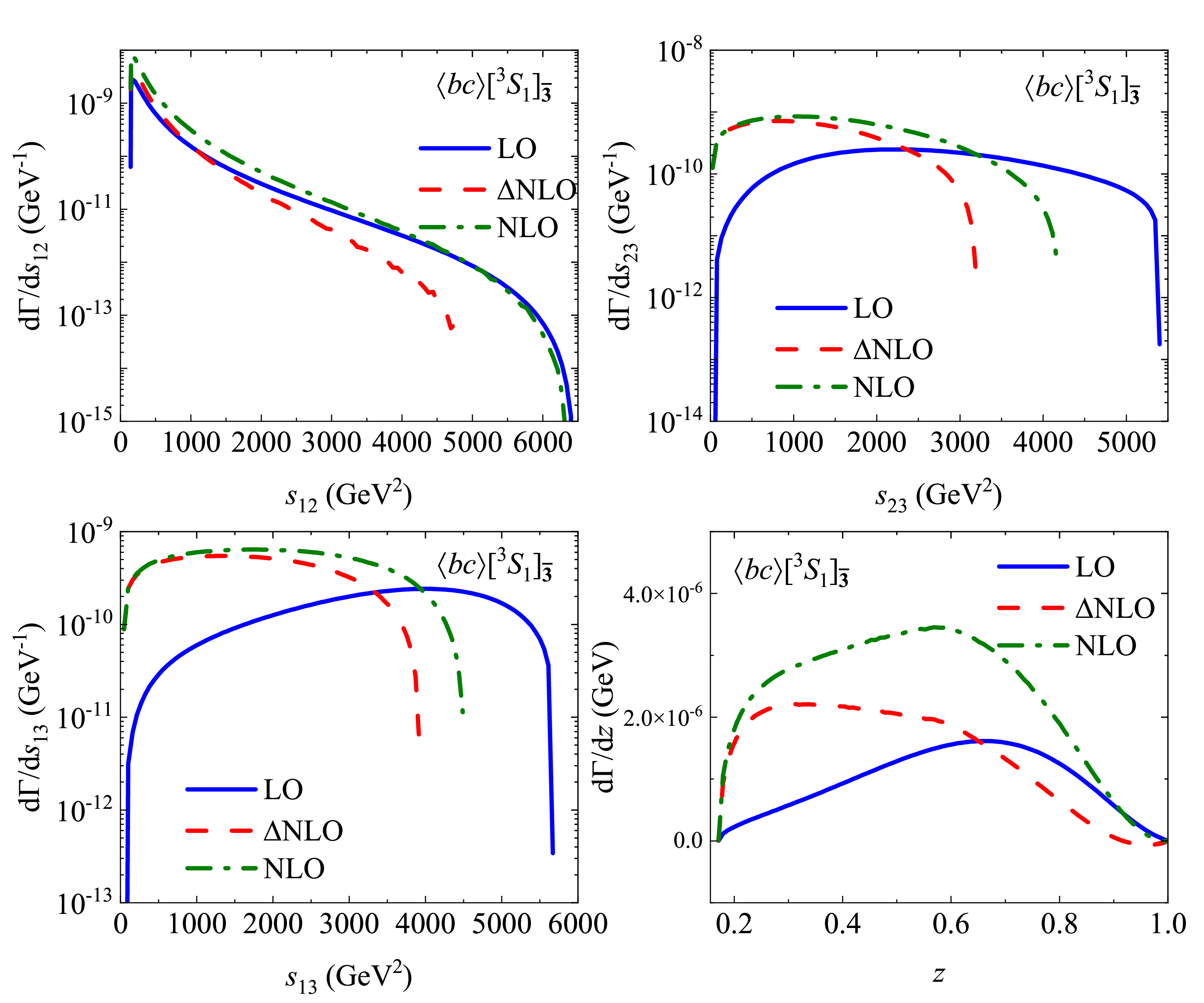}
\includegraphics[width=0.49\textwidth]{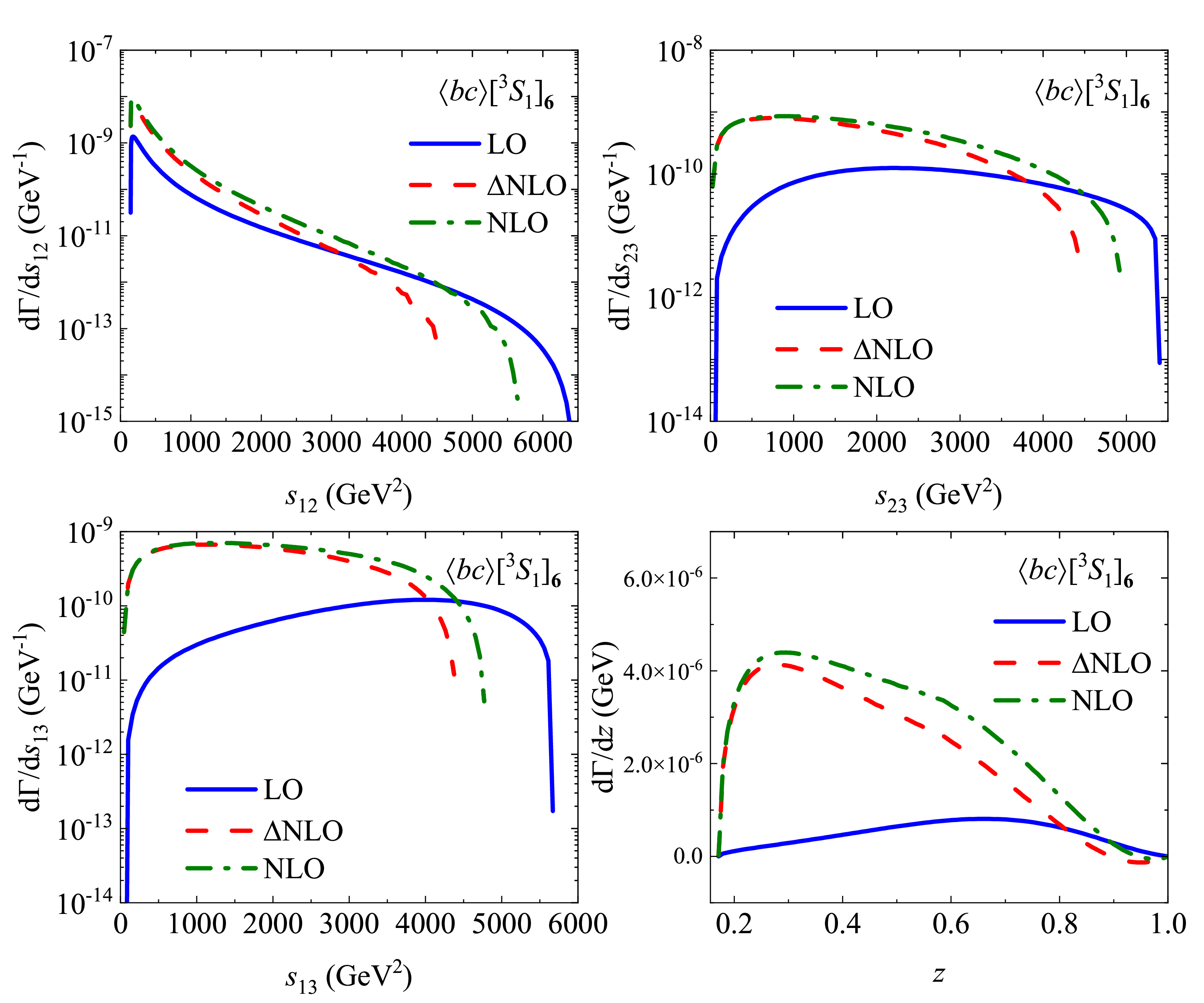}
\caption{Differential decay widths for $\langle bc\rangle[{}^{3}S_{1}]_{\bar{\mathbf{3}}}$ (left) and $\langle bc\rangle[{}^{3}S_{1}]_{\mathbf{6}}$ (right).}
\label{fig:Xibc_3s1_distributions}
\end{figure}

To examine the kinematic characteristics in detail, the differential decay widths are displayed in Figs.~\ref{fig:Xicc_distributions}--\ref{fig:Xibc_3s1_distributions}, including the invariant-mass and energy-fraction distributions. Here the invariant masses of the final-state pairs are defined as $s_{ij}=(p_{i}+p_{j})^{2}$, and the energy fraction as $z\equiv E_{1}/E_{1}^{\max}$, with $E_{1}^{\max}=(m_{W}^{2}+M^{2}-m_{Q}^{2})/(2m_{W})$ evaluated in the $W$ rest frame, the minimal invariant mass of the recoil system being $m_{Q}$. The blue solid curve is the LO result, the red dashed curve labelled ``$\Delta$NLO'' is the NLO correction $\mathrm{d}\Delta\Gamma_{\mathrm{NLO}}/\mathrm{d}x$, and the green dash-dotted curve labelled ``NLO'' is the total NLO result, $\mathrm{d}\Gamma_{\mathrm{tot}}/\mathrm{d}x$.

At NLO, observables involving the massless $\bar{s}$, such as $s_{13}$ and $s_{23}$, must be collinear safe; i.e., in the limit where the $\bar{s}$ and the emitted gluon become collinear, they must depend on the combined momentum $p_{3}+k$ rather than on the bare $p_{3}$~\cite{Catani:1996vz}. This issue does not affect the collinear safety of the inclusive width, nor of the $s_{12}$ and $z$ distributions.

The $s_{12}$ distributions in Figs.~\ref{fig:Xicc_distributions}--\ref{fig:Xibc_3s1_distributions} are concentrated near the heavy-pair threshold $(M+m_{Q})^{2}$ and decrease rapidly as $s_{12}$ increases. The NLO correction is positive in the low-$s_{12}$ region but turns negative at larger $s_{12}$, where the total NLO result can fall below the LO one. In the $s_{13}$ and $s_{23}$ distributions of Figs.~\ref{fig:Xicc_distributions}--\ref{fig:Xibc_3s1_distributions}, the LO maxima occur at intermediate invariant masses, whereas the NLO correction is weighted more strongly toward the low- and intermediate-mass regions, so that the total NLO maxima are shifted to smaller invariant masses.

For $\Xi_{cc}$ production, the integrated corrections are $1.96\,\Gamma_{\mathrm{LO}}$ for $\langle cc\rangle[{}^{1}S_{0}]_{\mathbf{6}}$ and $0.94\,\Gamma_{\mathrm{LO}}$ for $\langle cc\rangle[{}^{3}S_{1}]_{\bar{\mathbf{3}}}$. This difference is also visible in the $z$ distributions of Fig.~\ref{fig:Xicc_distributions}, mainly in the low- and intermediate-$z$ region: the sextet spectrum is predominantly generated by the NLO correction and is $1.96$ times its LO value, whereas the antitriplet spectrum retains a larger LO component. In both cases, the total NLO distribution reaches its maximum at a smaller $z$ than the LO curve and is suppressed near the upper endpoint. From Fig.~\ref{fig:Xibc_1s0_distributions}, one finds that, compared with the $\bar{\mathbf{3}}$ state, the NLO correction of the $\mathbf{6}$ state dominates over its LO contribution over most of the low-$z$ region. The spin-triplet comparison in Fig.~\ref{fig:Xibc_3s1_distributions} exhibits the strongest color dependence: for $\langle bc\rangle[{}^{3}S_{1}]_{\bar{\mathbf{3}}}$ the $z$ spectrum remains centered at an intermediate energy fraction, whereas for $\langle bc\rangle[{}^{3}S_{1}]_{\mathbf{6}}$ the NLO correction shifts the maximum toward substantially smaller $z$.

The $z$ spectra make the effect of real radiation most transparent. Real emission carries energy away from the diquark and enhances the small- and moderate-$z$ regions, while the LO contribution is concentrated at larger $z$. However, near $z\approx1$ the NLO correction becomes negative and can drive the total NLO result below zero, most visibly in the $\langle cc\rangle$ channels. Such negative values are artefacts of the fixed-order expansion rather than physical differential rates, and show that the endpoint region requires further treatment, such as the resummation of the large logarithms associated with the $z\to1$ limit.

\subsection{Discussion of theoretical uncertainties}

The theoretical uncertainties considered here arise from the heavy-quark masses and the renormalization scale. To estimate those induced by the heavy-quark masses, we vary $m_{c}=1.8\pm0.3~\mathrm{GeV}$ and $m_{b}=5.1\pm0.3~\mathrm{GeV}$ independently, with the corresponding results listed in Tables~\ref{tab:mc_uncertainty} and~\ref{tab:mb_uncertainty}. In these tables the sums are computed from the unrounded values and may therefore differ slightly from the sums of the displayed entries. Relative to the central result at $m_{c}=1.8~\mathrm{GeV}$, the total $\Xi_{cc}$ width increases by $85.35\%$ for $m_{c}=1.5~\mathrm{GeV}$ and decreases by $40.99\%$ for $m_{c}=2.1~\mathrm{GeV}$. $\Xi_{bc}$ depends on the heavy-quark masses much more weakly: the corresponding variations are $+10.80\%$ and $-7.17\%$ for $m_{c}$, and $+20.70\%$ and $-16.35\%$ for $m_{b}$.

\begin{table}[H]
\caption{Total NLO decay widths $\Gamma_{\mathrm{tot}}$ (in keV) for $W^{+}\to\Xi_{cc}+\bar{c}+\bar{s}+X$ and $W^{+}\to\Xi_{bc}+\bar{b}+\bar{s}+X$, with $m_{c}$ varied while $m_{b}=5.1~\mathrm{GeV}$ and $\mu_{R}$ fixed at its central value $2m_{c}=3.6~\mathrm{GeV}$. All masses are given in GeV.}
\label{tab:mc_uncertainty}
\centering
\begin{tabular*}{\textwidth}{@{\extracolsep{\fill}}lccc@{}}
\hline\hline
State & $m_{c}=1.5$ & $m_{c}=1.8$ & $m_{c}=2.1$ \\
\hline
$\langle cc\rangle[{}^{1}S_{0}]_{\mathbf{6}}$ & $143.57$ & $78.34$ & $46.62$ \\
$\langle cc\rangle[{}^{3}S_{1}]_{\bar{\mathbf{3}}}$ & $198.90$ & $106.43$ & $62.41$ \\
$\langle cc\rangle$-sum & $342.47$ & $184.77$ & $109.03$ \\
\hline
$\langle bc\rangle[{}^{1}S_{0}]_{\bar{\mathbf{3}}}$ & $2.05$ & $1.94$ & $1.87$ \\
$\langle bc\rangle[{}^{1}S_{0}]_{\mathbf{6}}$ & $1.60$ & $1.46$ & $1.37$ \\
$\langle bc\rangle[{}^{3}S_{1}]_{\bar{\mathbf{3}}}$ & $2.13$ & $1.95$ & $1.83$ \\
$\langle bc\rangle[{}^{3}S_{1}]_{\mathbf{6}}$ & $2.60$ & $2.20$ & $1.95$ \\
$\langle bc\rangle$-sum & $8.38$ & $7.56$ & $7.02$ \\
\hline\hline
\end{tabular*}
\end{table}

\begin{table}[H]
\caption{Total NLO decay widths $\Gamma_{\mathrm{tot}}$ (in keV) for $W^{+}\to\Xi_{bc}+\bar{b}+\bar{s}+X$, with $m_{b}$ varied while $m_{c}=1.8~\mathrm{GeV}$ and $\mu_{R}$ fixed at its central value $2m_{b}=10.2~\mathrm{GeV}$. All masses are given in GeV.}
\label{tab:mb_uncertainty}
\centering
\begin{tabular*}{\textwidth}{@{\extracolsep{\fill}}lccc@{}}
\hline\hline
State & $m_{b}=4.8$ & $m_{b}=5.1$ & $m_{b}=5.4$ \\
\hline
$\langle bc\rangle[{}^{1}S_{0}]_{\bar{\mathbf{3}}}$ & $2.37$ & $1.94$ & $1.61$ \\
$\langle bc\rangle[{}^{1}S_{0}]_{\mathbf{6}}$ & $1.76$ & $1.46$ & $1.22$ \\
$\langle bc\rangle[{}^{3}S_{1}]_{\bar{\mathbf{3}}}$ & $2.37$ & $1.95$ & $1.63$ \\
$\langle bc\rangle[{}^{3}S_{1}]_{\mathbf{6}}$ & $2.63$ & $2.20$ & $1.86$ \\
$\langle bc\rangle$-sum & $9.12$ & $7.56$ & $6.32$ \\
\hline\hline
\end{tabular*}
\end{table}

Table~\ref{tab:scale_uncertainty} presents the total NLO decay widths under the conventional scale variations $\mu_{R}\in[\mu_{0}/2, 2\mu_{0}]$, where $\mu_{0}=2m_{c}$ for $\Xi_{cc}$ and $\mu_{0}=2m_{b}$ for $\Xi_{bc}$. The corresponding strong couplings at the three scales $\mu_{0}/2$, $\mu_{0}$, and $2\mu_{0}$ are $(0.326,\,0.239,\,0.193)$ for $\Xi_{cc}$ and $(0.212,\,0.177,\,0.153)$ for $\Xi_{bc}$. Varying $\mu_{R}$ over this range, the total NLO decay widths become $\Gamma_{\mathrm{tot}}(\Xi_{cc})=184.77^{+76.74\%}_{-31.43\%}~\mathrm{keV}$ and $\Gamma_{\mathrm{tot}}(\Xi_{bc})=7.56^{+44.78\%}_{-25.41\%}~\mathrm{keV}$, where the upper and lower variations correspond to $\mu_{R}=\mu_{0}/2$ and $2\mu_{0}$, respectively. The residual scale uncertainty at NLO thus remains sizable, especially for $\Xi_{cc}$.

\begin{table}[H]
\caption{Total NLO decay widths $\Gamma_{\mathrm{tot}}$ (in keV) under the conventional scale variations $\mu_{R}\in[\mu_{0}/2, 2\mu_{0}]$, with $m_{c}=1.8~\mathrm{GeV}$ and $m_{b}=5.1~\mathrm{GeV}$.}
\label{tab:scale_uncertainty}
\centering
\begin{tabular*}{\textwidth}{@{\extracolsep{\fill}}lccc@{}}
\hline\hline
State & $\mu_{R}=\mu_{0}/2$ & $\mu_{R}=\mu_{0}$ & $\mu_{R}=2\mu_{0}$ \\
\hline
$\langle cc\rangle[{}^{1}S_{0}]_{\mathbf{6}}$ & $153.11$ & $78.34$ & $50.93$ \\
$\langle cc\rangle[{}^{3}S_{1}]_{\bar{\mathbf{3}}}$ & $173.45$ & $106.43$ & $75.77$ \\
$\langle cc\rangle$-sum & $326.56$ & $184.77$ & $126.70$ \\
\hline
$\langle bc\rangle[{}^{1}S_{0}]_{\bar{\mathbf{3}}}$ & $2.61$ & $1.94$ & $1.52$ \\
$\langle bc\rangle[{}^{1}S_{0}]_{\mathbf{6}}$ & $2.15$ & $1.46$ & $1.08$ \\
$\langle bc\rangle[{}^{3}S_{1}]_{\bar{\mathbf{3}}}$ & $2.72$ & $1.95$ & $1.50$ \\
$\langle bc\rangle[{}^{3}S_{1}]_{\mathbf{6}}$ & $3.47$ & $2.20$ & $1.54$ \\
$\langle bc\rangle$-sum & $10.94$ & $7.56$ & $5.64$ \\
\hline\hline
\end{tabular*}
\end{table}

\subsection{Scale-invariant predictions with the PMC}
\label{sec:pmc}

To eliminate the large ambiguity of the renormalization scale, we absorb the $\beta_{0}$-dependent NLO contribution into the running coupling using the single-scale PMC prescription~\cite{Brodsky:2011ta, Shen:2017pdu, Wu:2013ei, Wu:2019mky}. The NLO decay width can then be rewritten as
\begin{equation}
\Gamma_{\mathrm{tot}}^{\mathrm{PMC}}
=A\alpha_{s}^{2}(\mu_{\mathrm{PMC}})
\left[1+r_{\mathrm{conf}}\frac{\alpha_{s}(\mu_{\mathrm{PMC}})}{\pi}\right],
\end{equation}
where $A$ collects the LO electroweak factor and the long-distance matrix element (LDME), and $r_{\mathrm{conf}}$ is the conformal coefficient of the NLO correction. The PMC scales are $2.66$ and $2.84~\mathrm{GeV}$ for $\langle cc\rangle[{}^{1}S_{0}]_{\mathbf{6}}$ and $\langle cc\rangle[{}^{3}S_{1}]_{\bar{\mathbf{3}}}$, respectively. For the $\langle bc\rangle$ system they are $6.42~\mathrm{GeV}$ for both spin-singlet color states and $7.05~\mathrm{GeV}$ for the spin-triplet ones.

For the $\mathbf{6}$ and $\bar{\mathbf{3}}$ states of $\langle bc\rangle[{}^{3}S_{1}]$, the quark vacuum-polarization and coupling-renormalization terms carry the same overall color factor as the tree-level amplitude. This factor cancels in the normalized $\beta_{0}$ coefficient and hence gives the same PMC scale, although the conformal corrections and the resulting $K$ factors differ.

Table~\ref{tab:pmc_comparison} gives the PMC results, where $\Gamma_{\mathrm{LO}}^{\mathrm{PMC}}=A\alpha_{s}^{2}(\mu_{\mathrm{PMC}})$ denotes the LO piece evaluated at the PMC scale and $\Delta\Gamma_{\mathrm{NLO}}^{\mathrm{PMC}}=\Gamma_{\mathrm{tot}}^{\mathrm{PMC}} -\Gamma_{\mathrm{LO}}^{\mathrm{PMC}}$. The PMC total widths are $21.79\%$ ($\Xi_{cc}$) and $23.73\%$ ($\Xi_{bc}$) larger than the corresponding conventional NLO central values. Although PMC scale setting eliminates the initial renormalization-scale ambiguity at NLO, the conformal corrections remain large in several channels, which indicates that the convergence of the perturbative series is not yet established. A higher-order calculation is therefore required to assess that convergence and to obtain more reliable predictions.

\begin{table}[H]
\caption{Decay widths (in keV) and $K$ factors after PMC scale setting. The LO piece $\Gamma_{\mathrm{LO}}^{\mathrm{PMC}}$ is evaluated at the PMC scale $\mu_{\mathrm{PMC}}$.}
\label{tab:pmc_comparison}
\centering
\begin{tabular*}{\textwidth}{@{\extracolsep{\fill}}lcccc@{}}
\hline\hline
State & $\Gamma_{\mathrm{LO}}^{\mathrm{PMC}}$ & $\Delta\Gamma_{\mathrm{NLO}}^{\mathrm{PMC}}$ & $\Gamma_{\mathrm{tot}}^{\mathrm{PMC}}$ & $K$ \\
\hline
$\langle cc\rangle[{}^{1}S_{0}]_{\mathbf{6}}$ & $33.83$ & $67.41$ & $101.24$ & $2.99$ \\
$\langle cc\rangle[{}^{3}S_{1}]_{\bar{\mathbf{3}}}$ & $66.28$ & $57.51$ & $123.79$ & $1.87$ \\
$\langle cc\rangle$-sum & $100.11$ & $124.92$ & $225.04$ & $2.25$ \\
\hline
$\langle bc\rangle[{}^{1}S_{0}]_{\bar{\mathbf{3}}}$ & $1.15$ & $1.21$ & $2.37$ & $2.05$ \\
$\langle bc\rangle[{}^{1}S_{0}]_{\mathbf{6}}$ & $0.58$ & $1.30$ & $1.88$ & $3.26$ \\
$\langle bc\rangle[{}^{3}S_{1}]_{\bar{\mathbf{3}}}$ & $0.96$ & $1.36$ & $2.32$ & $2.42$ \\
$\langle bc\rangle[{}^{3}S_{1}]_{\mathbf{6}}$ & $0.48$ & $2.30$ & $2.78$ & $5.79$ \\
$\langle bc\rangle$-sum & $3.17$ & $6.18$ & $9.35$ & $2.95$ \\
\hline\hline
\end{tabular*}
\end{table}

\section{Summary}
\label{sec:summary}

The $S$-wave short-distance coefficients for $\Xi_{cc}$ and $\Xi_{bc}$ production through $W^{+}$ decays have been evaluated at NLO in QCD. Defining $\mathcal{B}=\Gamma_{\mathrm{tot}}/\Gamma_{W}$ with $\Gamma_{W}=2.085~\mathrm{GeV}$, the inclusive branching ratios are $\mathcal{B}(W^{+}\to\Xi_{cc}+\bar{c}+\bar{s}+X)\approx8.86\times10^{-5}$ and $\mathcal{B}(W^{+}\to\Xi_{bc}+\bar{b}+\bar{s}+X)\approx3.63\times10^{-6}$. With the reference quark masses, the summed $K$ factors are $2.27$ and $2.95$ for $\Xi_{cc}$ and $\Xi_{bc}$ production, respectively. The correction is particularly large for $\langle bc\rangle[{}^{3}S_{1}]_{\mathbf{6}}$, where the $K$ factor reaches $5.53$; this suggests that a higher-order calculation is needed before a precision estimate can be assigned. Under the conventional scale variation $\mu_{R}\in[\mu_{0}/2,2\mu_{0}]$, with $\mu_{0}=2m_{c}$ ($2m_{b}$) for $\Xi_{cc}$ ($\Xi_{bc}$), the total NLO widths are \mbox{$\Gamma_{\mathrm{tot}}(\Xi_{cc})=184.77^{+76.74\%}_{-31.43\%}~\mathrm{keV}$} and \mbox{$\Gamma_{\mathrm{tot}}(\Xi_{bc})=7.56^{+44.78\%}_{-25.41\%}~\mathrm{keV}$}. After PMC scale setting, the $\mu_{R}$ ambiguity is eliminated and the widths increase to $225.04$ and $9.35~\mathrm{keV}$, respectively.

The soft contribution includes radiation off the colored diquark as well as its correlations with the two antiquarks; its color algebra and the resulting dipole integrals are collected in Appendix~\ref{app:soft-derivation}, with the final expressions also given in Sec.~\ref{sec:formalism}. The $s_{12}$ invariant-mass distribution is concentrated toward the heavy-pair threshold, while real radiation shifts the energy-fraction distribution toward smaller $z$. The heavy-quark-mass and renormalization-scale uncertainties have been estimated separately and are not combined in quadrature. The predicted yields are also subject to the model dependence of the diquark-to-baryon matrix elements, in particular the assumption $h_{\mathbf{6}}=h_{\bar{\mathbf{3}}}$. Consequently, these results are best regarded as inclusive benchmark predictions rather than detector-level discovery yields.

\acknowledgments
This work was supported by the Natural Science Foundation of Guangxi (No. 2025GXNSFAA069775) and by the National Natural Science Foundation of China (Grants No. 12505106, 12665014, 12547101). H.-H. Ma is also supported by the S\~ao Paulo Research Foundation (FAPESP), Brasil, grant number 2025/01276-7. X.-G. Wu and X.-C. Zheng are also supported by the Fundamental Research Funds for the Central Universities Grant No. 2026CDJZKPT-10.

\appendix
\section{Color correlations and soft dipole integrals}
\label{app:soft-derivation}

\subsection{Color correlations}

At leading order in the relative velocity, $p_Q=(m_{Q}/M)p_1$ and $p_c=(m_{c}/M)p_1$, with $M=m_{Q}+m_{c}$. The two eikonal currents therefore combine into one current with color charge $\mathbf{T}_1=\mathbf{T}_Q+\mathbf{T}_c$. Here $\mathbf{T}_i$ is the color matrix of the particle carrying momentum $p_{i}$. Taking all three hard legs as outgoing, color conservation at LO amplitude gives
\begin{equation} \label{color:conservation}
 (\mathbf{T}_1+\mathbf{T}_2+\mathbf{T}_3)|\mathcal{M}_{\mathrm{LO}}^{R}\rangle=0,
 \qquad \mathbf{T}_1^2=C_{R},\quad \mathbf{T}_2^2=\mathbf{T}_3^2=C_{F},
\end{equation}
where $C_{\bar{\mathbf{3}}}=4/3$, $C_{\mathbf{6}}=10/3$, and $C_{F}=4/3$. The normalized color correlation is
\begin{equation} 
 \mathcal{C}_{ij}^{(R)}=
 \frac{\langle\mathcal{M}_{\mathrm{LO}}^{R}|\mathbf{T}_i\cdot\mathbf{T}_j|
 \mathcal{M}_{\mathrm{LO}}^{R}\rangle}
 {\langle\mathcal{M}_{\mathrm{LO}}^{R}|\mathcal{M}_{\mathrm{LO}}^{R}\rangle}.
\end{equation}
By using the color conservation relation shown in~Eq. \eqref{color:conservation}, we can obtain
\begin{align}
 \mathcal{C}_{11}^{(R)}&=C_{R},\qquad
 \mathcal{C}_{22}^{(R)}=\mathcal{C}_{33}^{(R)}=C_{F},\nonumber\\
 \mathcal{C}_{12}^{(R)}&=\mathcal{C}_{13}^{(R)}=-\frac{C_{R}}{2},\qquad
 \mathcal{C}_{23}^{(R)}=\frac{C_{R}-2C_{F}}{2}.
 \label{eq:soft-color-correlations}
\end{align}
The soft amplitude is a sum of emissions from the three hard legs~\cite{Catani:1996vz}:
\begin{equation}
 |\mathcal{M}_{S}^{a}\rangle=g_s\mu_{R}^{\varepsilon}\varepsilon_{\mu}^{*}(k)
 \sum_{i=1}^{3}T_{i}^{a}\frac{p_{i}^\mu}{p_{i}\cdot k}
 |\mathcal{M}_{\mathrm{LO}}^{R}\rangle.
\end{equation}
Squaring this sum pairs the emission from leg $i$ with that from leg $j$. Using color conservation in the gluon-polarization sum gives
\begin{equation}
 \sum|\mathcal{M}_{\mathrm{Real}}^{\mathrm{soft}}|^{2}
 =-g_s^2\mu_{R}^{2\varepsilon}\sum_{i,j=1}^3
 \mathcal{C}_{ij}^{(R)}\frac{p_{i}\cdot p_{j}}{(p_{i}\cdot k)(p_{j}\cdot k)}
 \sum|\mathcal{M}_{\mathrm{LO}}^{R}|^{2}.
 \label{eq:soft-eikonal-main}
\end{equation}
The terms with $i=j$ are self-correlations; those with $i\ne j$ are interference terms between different outgoing legs.

\subsection{Integral convention}

We use $D=4-2\varepsilon$, $k=p_{4}$, $p_1^2=M^2$, $p_2^2=m_{Q}^2$, and $p_{3}^2=k^2=0$. Here $p_{i}\cdot p_{j}=E_{i}E_j-\mathbf{p}_i\cdot\mathbf{p}_j$ is a four-momentum scalar product. All energies and angles are evaluated in the $W$ rest frame. Following the soft energy slicing of Ref.~\cite{Harris:2001sx}, we integrate over $0<E_{g}<E_{\max}=m_{W}\delta_{s}/2$ and define
\begin{align}
 &4\pi\alpha_{s}\mu_{R}^{2\varepsilon}
 \int_{E_{g}<E_{\max}}\frac{\mathrm{d}^{D-1}\mathbf{k}}{2E_{g}(2\pi)^{D-1}}
 \frac{p_{i}\cdot p_{j}}{(p_{i}\cdot k)(p_{j}\cdot k)}\nonumber\\
 &\hspace{3em}=\frac{\alpha_{s}}{2\pi}
 \frac{\Gamma(1-\varepsilon)}{\Gamma(1-2\varepsilon)}
 \left(\frac{4\pi\mu_{R}^2}{m_{W}^2}\right)^\varepsilon I_{ij}.
 \label{eq:app-Iij-master}
\end{align}
Integrating Eq.~\eqref{eq:soft-eikonal-main} with this definition and extracting the common $C_{F}$ in Eq.~\eqref{eq:soft-width} gives
\begin{align}
 S_R&=-\frac1{C_{F}}\sum_{i,j=1}^{3}\mathcal{C}_{ij}^{(R)}I_{ij}\nonumber\\
 &=-\frac{C_{R}}{C_{F}}I_{11}-I_{22}
   +\frac{C_{R}}{C_{F}}(I_{12}+I_{13})
   +\left(2-\frac{C_{R}}{C_{F}}\right)I_{23}.
 \label{eq:app-S-dipoles}
\end{align}

Here the relations $I_{ji}=I_{ij}$ and $I_{33}=0$ are already applied. Substituting $C_{R}=4/3$ or $10/3$ for $\bar{\mathbf{3}}$ or $\mathbf{6}$ respectively, gives us the two expressions in Eq.~\eqref{eq:soft-unified}.

At leading soft order, the phase space can be factorized as $\mathrm{d}\Phi_{4}|_S=\mathrm{d}\Phi_{3}\,\mathrm{d}^{D-1}\mathbf{k}/\allowbreak [2E_{g}(2\pi)^{D-1}]$. The energy integral is
\begin{equation}
 \int_0^{E_{\max}}\mathrm{d}E_{g}\,E_{g}^{-1-2\varepsilon}
 =-\frac{E_{\max}^{-2\varepsilon}}{2\varepsilon}.
\end{equation}
The angular integrals are expanded through $O(\varepsilon)$ using Appendix B of Ref.~\cite{Harris:2001sx}, retaining the finite terms after energy integration. 

\subsection{Explicit integrals}

Let
\begin{equation}
 L_{s}=\ln\delta_{s},\qquad
 \beta_{i}=\frac{|\mathbf{p}_i|}{E_{i}},\qquad
 L_{i}=\ln\frac{1+\beta_{i}}{1-\beta_{i}},\qquad i=1,2.
\end{equation}
The three diagonal integrals are
\begin{align}
 I_{11}&=-\frac1\varepsilon+2L_{s}-\frac{L_1}{\beta_1}+O(\varepsilon),\nonumber\\
 I_{22}&=-\frac1\varepsilon+2L_{s}-\frac{L_2}{\beta_2}+O(\varepsilon),\nonumber\\
 I_{33}&=0.
 \label{eq:app-Iii}
\end{align}
The ratios $L_{i}/\beta_{i}$ tend to $2$ as $\beta_{i}\to0$.

For the massive--massless pairs, $\theta$ is the angle between the spatial momenta. Define
\begin{align}
 \ell(\beta,\theta)&=\ln\frac{(1-\beta\cos\theta)^2}{1-\beta^2},\nonumber\\
 H(\beta,\theta)&=\ln^2\!\left(\frac{1-\beta}{1-\beta\cos\theta}\right)
 -\frac12\ln^2\!\left(\frac{1+\beta}{1-\beta}\right)\nonumber\\
 &\quad+2\operatorname{Li}_2\!\left[-\frac{\beta(1-\cos\theta)}{1-\beta}\right]
 -2\operatorname{Li}_2\!\left[-\frac{\beta(1+\cos\theta)}{1-\beta\cos\theta}\right].
 \label{eq:app-H}
\end{align}
These functions are real for $0\le\beta<1$ and $0\le\theta\le\pi$. Defining $\cos\theta_{i3}=\hat{\mathbf{p}}_i\cdot\hat{\mathbf{p}}_3$, $\ell_{i3}=\ell(\beta_{i},\theta_{i3})$, and $H_{i3}=H(\beta_{i},\theta_{i3})$, we obtain
\begin{align}
 I_{13}&=\frac1{2\varepsilon^{2}}-\frac{L_{s}+\ell_{13}/2}{\varepsilon}
       +L_{s}^2+L_{s}\ell_{13}+\frac{H_{13}}2+O(\varepsilon),\nonumber\\
 I_{23}&=\frac1{2\varepsilon^{2}}-\frac{L_{s}+\ell_{23}/2}{\varepsilon}
       +L_{s}^2+L_{s}\ell_{23}+\frac{H_{23}}2+O(\varepsilon).
 \label{eq:app-Ipq}
\end{align}

For the massive--massive pair, we need to introduce the following additional definitions.

\begin{align}
 u_{12}&=p_1\cdot p_2,&
 \kappa_{12}&=\sqrt{u_{12}^2-M^2m_{Q}^2},\nonumber\\
 R_{12}&=\frac{u_{12}}{\kappa_{12}},&
 \lambda_{12}&=\ln\frac{u_{12}+\kappa_{12}}{Mm_{Q}},\nonumber\\
 \alpha_{12}&=\frac{u_{12}+\kappa_{12}}{M^2},&
 q_{12}&=\alpha_{12}p_1-p_2.
\end{align}

With the similar definitions $\cos\theta_{iq}=\hat{\mathbf{p}}_i\cdot\hat{\mathbf q}_{12}$, $H_{qi}=H(\beta_{i},\theta_{iq})$, and $\ell_{iq}=\ell(\beta_{i},\theta_{iq})$, and the same energy cutoff, the eikonal identity gives $I_{12}=R_{12}(I_{q2}-I_{q1})$. Using $\ell_{2q}-\ell_{1q}=2\lambda_{12}$, we find

\begin{equation}
 I_{12}=R_{12}\left[-\frac{\lambda_{12}}\varepsilon
 +2\lambda_{12}L_{s}+\frac{H_{q2}-H_{q1}}2\right]+O(\varepsilon).
 \label{eq:app-I12}
\end{equation}


\begin{thebibliography}{99}
\interlinepenalty=10000

\bibitem{Caswell:1985ui}
W.~E.~Caswell and G.~P.~Lepage,
Effective Lagrangians for bound state problems in QED, QCD, and other field theories,
Phys.\ Lett.\ B \textbf{167}, 437 (1986).

\bibitem{Bodwin:1994jh}
G.~T.~Bodwin, E.~Braaten, and G.~P.~Lepage,
Rigorous QCD analysis of inclusive annihilation and production of heavy quarkonium,
Phys.\ Rev.\ D \textbf{51}, 1125 (1995); \textbf{55}, 5853(E) (1997).

\bibitem{Brambilla:2004jw}
N.~Brambilla, A.~Pineda, J.~Soto, and A.~Vairo,
Effective field theories for heavy quarkonium,
Rev.\ Mod.\ Phys.\ \textbf{77}, 1423 (2005).

\bibitem{Ma:2003zk}
J.~P.~Ma and Z.~G.~Si,
Factorization approach for inclusive production of doubly heavy baryon,
Phys.\ Lett.\ B \textbf{568}, 135 (2003).

\bibitem{Fleming:2005pd}
S.~Fleming and T.~Mehen,
Doubly heavy baryons, heavy quark--diquark symmetry and NRQCD,
Phys.\ Rev.\ D \textbf{73}, 034502 (2006).

\bibitem{Cohen:2006jg}
T.~D.~Cohen and P.~M.~Hohler,
Doubly heavy hadrons and the domain of validity of doubly heavy diquark--antiquark symmetry,
Phys.\ Rev.\ D \textbf{74}, 094003 (2006).

\bibitem{Baranov:1995rc}
S.~P.~Baranov,
On the production of doubly flavored baryons in $pp$, $ep$, and $\gamma\gamma$ collisions,
Phys.\ Rev.\ D \textbf{54}, 3228 (1996).

\bibitem{Chang:2006eu}
C.~H.~Chang, C.~F.~Qiao, J.~X.~Wang and X.~G.~Wu,
Estimate of the hadronic production of the doubly charmed baryon Xi(cc) under GM-VFN scheme,
Phys.\ Rev.\ D \textbf{73}, 094022 (2006).

\bibitem{Chang:2006xp}
C.~H.~Chang, J.~P.~Ma, C.~F.~Qiao and X.~G.~Wu,
Hadronic production of the doubly charmed baryon Xi(cc) with intrinsic charm,
J.\ Phys.\ G \textbf{34}, 845 (2007).

\bibitem{Chang:2007pp}
C.~H.~Chang, J.~X.~Wang and X.~G.~Wu,
GENXICC: A Generator for hadronic production of the double heavy baryons Xi(cc), Xi(bc) and Xi(bb),
Comput.\ Phys.\ Commun. \textbf{177}, 467 (2007).

\bibitem{Chen:2018koh}
G.~Chen, C.~H.~Chang and X.~G.~Wu,
Hadronic production of the doubly charmed baryon via the proton{\textendash}nucleus and the nucleus{\textendash}nucleus collisions at the RHIC and LHC,
Eur.\ Phys.\ J.\ C \textbf{78}, 801 (2018).

\bibitem{Chen:2019ykv}
G.~Chen, X.~G.~Wu and S.~Xu,
Impacts of the intrinsic charm content of the proton on the $\Xi_{cc}$ hadroproduction at a fixed target experiment at the LHC,
Phys.\ Rev.\ D \textbf{100}, 054022 (2019).

\bibitem{Zhang:2011hi}
J.~W.~Zhang, X.~G.~Wu, T.~Zhong, Y.~Yu, and Z.~Y.~Fang,
Hadronic production of the doubly heavy baryon $\Xi_{bc}$ at the LHC,
Phys.\ Rev.\ D \textbf{83}, 034026 (2011).

\bibitem{Chen:2014frw}
G.~Chen, X.~G.~Wu, J.~W.~Zhang, H.~Y.~Han, and H.~B.~Fu,
Hadronic production of $\Xi_{cc}$ at a fixed-target experiment at the LHC,
Phys.\ Rev.\ D \textbf{89}, 074020 (2014).

\bibitem{Jiang:2012jt}
J.~Jiang, X.~G.~Wu, Q.~L.~Liao, X.~C.~Zheng and Z.~Y.~Fang,
Doubly Heavy Baryon Production at A High Luminosity $e^+ e^-$ Collider,
Phys.\ Rev.\ D \textbf{86}, 054021 (2012).

\bibitem{Jiang:2013ej}
J.~Jiang, X.-G.~Wu, S.-M.~Wang, J.-W.~Zhang, and Z.-Y.~Fang,
Further study on the doubly heavy baryon production around the $Z^0$ peak at a high-luminosity $e^+e^-$ collider,
Phys.\ Rev.\ D \textbf{87}, 054027 (2013).

\bibitem{Zheng:2015ixa}
X.~C.~Zheng, C.~H.~Chang, and Z.~Pan,
Production of doubly heavy-flavored hadrons at $e^+e^-$ colliders,
Phys.\ Rev.\ D \textbf{93}, 034019 (2016).

\bibitem{Niu:2023ojf}
J.-J.~Niu, J.-B.~Li, H.-Y.~Bi, and H.-H.~Ma,
Production of excited doubly heavy baryons at the super-$Z$ factory,
Eur.\ Phys.\ J.\ C \textbf{83}, 822 (2023).

\bibitem{Li:2026abl}
H.~T.~Li, X.~C.~Zheng, X.~G.~Wu, J.~Yan and Z.~Yang,
EEXICC: An event generator for doubly heavy baryon production at $e^+e^-$ colliders,
Comput.\ Phys.\ Commun. \textbf{330}, 110461 (2027).

\bibitem{Li:2007vy}
S.~Y.~Li, Z.~G.~Si and Z.~J.~Yang,
Doubly heavy baryon production at gamma gamma collider,
Phys.\ Lett.\ B \textbf{648}, 284 (2007).

\bibitem{Chen:2014hqa}
G.~Chen, X.-G.~Wu, Z.~Sun, Y.~Ma, and H.-B.~Fu,
Photoproduction of doubly heavy baryon at the ILC,
JHEP \textbf{12}, 018 (2014).

\bibitem{Bi:2016vbt}
H.~Y.~Bi, R.~Y.~Zhang, X.~G.~Wu, W.~G.~Ma, X.~Z.~Li, and S.~Owusu,
Photoproduction of doubly heavy baryon at the LHeC,
Phys.\ Rev.\ D \textbf{95}, 074020 (2017).

\bibitem{Sun:2020mvl}
Z.~Sun and X.-G.~Wu,
Production of the doubly charmed baryon in deeply inelastic $ep$ scattering at the Large Hadron Electron Collider,
JHEP \textbf{07}, 034 (2020).

\bibitem{Zhan:2023fhh}
X.-J.~Zhan, X.-G.~Wu, and X.-C.~Zheng,
Photoproduction of $P$-wave doubly charmed baryon at future $e^+e^-$ collider,
JHEP \textbf{11}, 029 (2023).

\bibitem{Yang:2024ysg}
H.~Yang, J.~Jiang and B.~Long,
Doubly heavy hadron production in ultraperipheral collision,
Phys.\ Rev.\ D \textbf{109}, 114034 (2024).

\bibitem{Niu:2018ycb}
J.~J.~Niu, L.~Guo, H.~H.~Ma, X.~G.~Wu, and X.~C.~Zheng,
Production of semi-inclusive doubly heavy baryons via top-quark decays,
Phys.\ Rev.\ D \textbf{98}, 094021 (2018).

\bibitem{Ma:2022cgt}
H.-H.~Ma, J.-J.~Niu, and X.-C.~Zheng,
Excited doubly heavy baryons production via top-quark decays,
Phys.\ Rev.\ D \textbf{107}, 014006 (2023).

\bibitem{Niu:2019xuq}
J.~J.~Niu, L.~Guo, H.~H.~Ma, and X.~G.~Wu,
Production of doubly heavy baryons via Higgs boson decays,
Eur.\ Phys.\ J.\ C \textbf{79}, 339 (2019).

\bibitem{Ma:2022ger}
H.-H.~Ma and J.-J.~Niu,
Excited doubly heavy baryons production via Higgs decays,
Eur.\ Phys.\ J.\ C \textbf{83}, 5 (2023).

\bibitem{Zhang:2022jst}
P.~H.~Zhang, L.~Guo, X.~C.~Zheng, and Q.~W.~Ke,
Excited doubly heavy baryon production via $W^{+}$ boson decays,
Phys.\ Rev.\ D \textbf{105}, 034016 (2022).

\bibitem{Luo:2022jxq}
X.~Luo, H.-B.~Fu, and H.-J.~Tian,
Investigation of $Z$-boson decay into $\Xi_{bc}$ and $\Xi_{bb}$ baryons within the NRQCD factorization approach,
Chin.\ Phys.\ C \textbf{47}, 053102 (2023).

\bibitem{Tian:2023yhs}
H.-J.~Tian, X.~Luo, and H.-B.~Fu,
Further study on the production of $P$-wave doubly heavy baryons from $Z$-boson decays,
Phys.\ Lett.\ B \textbf{847}, 138302 (2023).

\bibitem{LHCb:2017iph}
R.~Aaij \textit{et al.} (LHCb Collaboration),
Observation of the doubly charmed baryon $\Xi_{cc}^{++}$,
Phys.\ Rev.\ Lett.\ \textbf{119}, 112001 (2017).

\bibitem{LHCb:2018zpl}
R.~Aaij \textit{et al.} (LHCb Collaboration),
Measurement of the lifetime of the doubly charmed baryon $\Xi_{cc}^{++}$,
Phys.\ Rev.\ Lett.\ \textbf{121}, 052002 (2018).

\bibitem{LHCb:2019qed}
R.~Aaij \textit{et al.} (LHCb Collaboration),
Precision measurement of the $\Xi_{cc}^{++}$ mass,
JHEP \textbf{02}, 049 (2020).

\bibitem{LHCb:2026xicc}
R.~Aaij \textit{et al.} (LHCb Collaboration),
Observation of the doubly charmed baryon $\Xi_{cc}^{+}$ with the LHCb Run 3 detector,
Phys.\ Rev.\ Lett.\ \textbf{137}, 021902 (2026).

\bibitem{LHCb:2020iko}
R.~Aaij \textit{et al.} (LHCb Collaboration),
Search for the doubly heavy $\Xi_{bc}^{0}$ baryon via decays to $D^0pK^-$,
JHEP \textbf{11}, 095 (2020).

\bibitem{LHCb:2021xba}
R.~Aaij \textit{et al.} (LHCb Collaboration),
Search for the doubly heavy baryons $\Omega_{bc}^{0}$ and $\Xi_{bc}^{0}$ decaying to $\Lambda_c^+\pi^-$ and $\Xi_c^+\pi^-$,
Chin.\ Phys.\ C \textbf{45}, 093002 (2021).

\bibitem{LHCb:2022vqr}
R.~Aaij \textit{et al.} (LHCb Collaboration),
Search for the doubly heavy baryon $\Xi_{bc}^{+}$ decaying to $J/\psi\,\Xi_c^{+}$,
Chin.\ Phys.\ C \textbf{47}, 093001 (2023).

\bibitem{Kiselev:2001fw}
V.~V.~Kiselev and A.~K.~Likhoded,
Baryons with two heavy quarks,
Phys.\ Usp.\ \textbf{45}, 455 (2002).

\bibitem{Gershtein:2000nx}
S.~S.~Gershtein, V.~V.~Kiselev, A.~K.~Likhoded, and A.~I.~Onishchenko,
Spectroscopy of doubly heavy baryons,
Phys.\ Rev.\ D \textbf{62}, 054021 (2000).

\bibitem{Ebert:2002ig}
D.~Ebert, R.~N.~Faustov, V.~O.~Galkin, and A.~P.~Martynenko,
Mass spectra of doubly heavy baryons in the relativistic quark model,
Phys.\ Rev.\ D \textbf{66}, 014008 (2002).

\bibitem{Karliner:2014gca}
M.~Karliner and J.~L.~Rosner,
Baryons with two heavy quarks: Masses, production, decays, and detection,
Phys.\ Rev.\ D \textbf{90}, 094007 (2014).

\bibitem{Padmanath:2015jea}
M.~Padmanath, R.~G.~Edwards, N.~Mathur, and M.~Peardon,
Spectroscopy of doubly-charmed baryons from lattice QCD,
Phys.\ Rev.\ D \textbf{91}, 094502 (2015).

\bibitem{Wang:2017mqp}
C.-Y.~Wang, C.~Meng, Y.-Q.~Ma, and K.-T.~Chao,
NLO effects for doubly heavy baryons in QCD sum rules,
Phys.\ Rev.\ D \textbf{99}, 014018 (2019).

\bibitem{Petrelli:1997ge}
A.~Petrelli, M.~Cacciari, M.~Greco, F.~Maltoni, and M.~L.~Mangano,
NLO production and decay of quarkonium,
Nucl.\ Phys.\ B \textbf{514}, 245 (1998).

\bibitem{Campbell:2007ws}
J.~Campbell, F.~Maltoni, and F.~Tramontano,
QCD corrections to $J/\psi$ and $\Upsilon$ production at hadron colliders,
Phys.\ Rev.\ Lett.\ \textbf{98}, 252002 (2007).

\bibitem{Butenschoen:2010rq}
M.~Butenschoen and B.~A.~Kniehl,
Reconciling $J/\psi$ production at HERA, RHIC, Tevatron, and LHC with NRQCD factorization at next-to-leading order,
Phys.\ Rev.\ Lett.\ \textbf{106}, 022003 (2011).

\bibitem{Ma:2010yw}
Y.-Q.~Ma, K.~Wang, and K.-T.~Chao,
$J/\psi(\psi')$ production at the Tevatron and LHC at $\mathcal{O}(\alpha_{s}^4v^4)$ in nonrelativistic QCD,
Phys.\ Rev.\ Lett.\ \textbf{106}, 042002 (2011).

\bibitem{Gong:2013qka}
B.~Gong, L.~P.~Wan, J.~X.~Wang and H.~F.~Zhang,
Complete next-to-leading-order study on the yield and polarization of $\Upsilon(1S,2S,3S)$ at the Tevatron and LHC,
Phys.\ Rev.\ Lett. \textbf{112}, 032001 (2014).

\bibitem{Sun:2010tx}
P.~Sun, L.~P.~Sun, and C.~F.~Qiao,
Next-to-leading-order corrections to top quark decays to heavy quarkonia,
Phys.\ Rev.\ D \textbf{81}, 114035 (2010).

\bibitem{Qiao:2011zj}
C.~F.~Qiao, L.~P.~Sun, and R.~L.~Zhu,
The NLO QCD corrections to $B_c$ meson production in $Z^0$ decays,
JHEP \textbf{08}, 131 (2011).

\bibitem{Jiang:2015pua}
J.~Jiang, L.-B.~Chen, and C.-F.~Qiao,
QCD NLO corrections to inclusive $B_c^*$ production in $Z^0$ decays,
Phys.\ Rev.\ D \textbf{91}, 034033 (2015).

\bibitem{Ke:2026BcNLO}
Q.-W.~Ke, L.~Guo, J.~Zeng, and X.-C.~Zheng,
Revisiting $B_c^{(*)}$ production in $Z$-boson decays at NLO in QCD,
arXiv:2610.21485 [hep-ph].

\bibitem{Zheng:2017xgj}
X.-C.~Zheng, C.-H.~Chang, T.-F.~Feng, and Z.~Pan,
NLO QCD corrections to $B_c(B_c^*)$ production around the $Z$ pole at an $e^+e^-$ collider,
Sci.\ China Phys.\ Mech.\ Astron.\ \textbf{61}, 031012 (2018).

\bibitem{Berezhnoy:2016hld}
A.~V.~Berezhnoy, A.~K.~Likhoded, A.~I.~Onishchenko, and S.~V.~Poslavsky,
Next-to-leading-order QCD corrections to paired $B_c$ production in $e^+e^-$ annihilation,
Nucl.\ Phys.\ B \textbf{915}, 224 (2017).

\bibitem{Chen:2019hck}
Z.-Q.~Chen, H.~Yang, and C.-F.~Qiao,
NLO QCD corrections to inclusive charmonium and $B_c$ meson production in $W^{+}$ decays,
Phys.\ Rev.\ D \textbf{101}, 036009 (2020).


\bibitem{Zheng:2019gnb}
X.~C.~Zheng, C.~H.~Chang, X.~G.~Wu, J.~Zeng, and X.~D.~Huang,
Next-to-leading-order QCD corrections to the production of $B_c$ and $B_c^*$ through $W^{+}$-boson decays,
Phys.\ Rev.\ D \textbf{101}, 034029 (2020).

\bibitem{Qiao:2011yk}
C.~F.~Qiao, L.~P.~Sun, D.~S.~Yang and R.~L.~Zhu,
``W Boson Inclusive Decays to Quarkonium at the LHC,''
Eur.\ Phys.\ J.\ C \textbf{71}, 1766 (2011).

\bibitem{Brodsky:2011ta}
S.~J.~Brodsky and X.~G.~Wu,
Scale setting using the extended renormalization group and the Principle of Maximum Conformality: The QCD coupling constant at four loops,
Phys.\ Rev.\ D \textbf{85}, 034038 (2012).

\bibitem{Mojaza:2012mf}
M.~Mojaza, S.~J.~Brodsky and X.~G.~Wu,
Systematic All-Orders Method to Eliminate Renormalization-Scale and Scheme Ambiguities in Perturbative QCD,
Phys.\ Rev.\ Lett. \textbf{110}, 192001 (2013).

\bibitem{Brodsky:2013vpa}
S.~J.~Brodsky, M.~Mojaza and X.~G.~Wu,
Systematic Scale-Setting to All Orders: The Principle of Maximum Conformality and Commensurate Scale Relations,
Phys.\ Rev.\ D \textbf{89}, 014027 (2014).

\bibitem{Shen:2017pdu}
J.-M.~Shen, X.-G.~Wu, B.-L.~Du, and S.~J.~Brodsky,
Novel all-orders single-scale approach to QCD renormalization scale-setting,
Phys.\ Rev.\ D \textbf{95}, 094006 (2017).

\bibitem{Wu:2013ei}
X.~G.~Wu, S.~J.~Brodsky, and M.~Mojaza,
The renormalization scale-setting problem in QCD,
Prog.\ Part.\ Nucl.\ Phys.\ \textbf{72}, 44 (2013).

\bibitem{Wu:2014iba}
X.~G.~Wu, Y.~Ma, S.~Q.~Wang, H.~B.~Fu, H.~H.~Ma, S.~J.~Brodsky and M.~Mojaza,
Renormalization Group Invariance and Optimal QCD Renormalization Scale-Setting,
Rept.\ Prog.\ Phys. \textbf{78}, 126201 (2015).

\bibitem{Wu:2019mky}
X.~G.~Wu, J.~M.~Shen, B.~L.~Du, X.~D.~Huang, S.~Q.~Wang, and S.~J.~Brodsky,
The QCD renormalization group equation and the elimination of fixed-order scheme-and-scale ambiguities using the Principle of Maximum Conformality,
Prog.\ Part.\ Nucl.\ Phys.\ \textbf{108}, 103706 (2019).

\bibitem{DiGiustino:2023jiq}
L.~Di Giustino, S.~J.~Brodsky, P.~G.~Ratcliffe, X.~G.~Wu and S.~Q.~Wang,
High precision tests of QCD without scale or scheme ambiguities: The 40th anniversary of the Brodsky{\textendash}Lepage{\textendash}Mackenzie method,
Prog.\ Part.\ Nucl.\ Phys. \textbf{135}, 104092 (2024).

\bibitem{Hahn:2000kx}
T.~Hahn,
Generating Feynman diagrams and amplitudes with FeynArts 3,
Comput.\ Phys.\ Commun.\ \textbf{140}, 418 (2001).

\bibitem{Mertig:1990an}
R.~Mertig, M.~B\"ohm, and A.~Denner,
Feyn Calc -- Computer-algebraic calculation of Feynman amplitudes,
Comput.\ Phys.\ Commun.\ \textbf{64}, 345 (1991).

\bibitem{Harris:2001sx}
B.~W.~Harris and J.~F.~Owens,
The two-cutoff phase-space slicing method,
Phys.\ Rev.\ D \textbf{65}, 094032 (2002).

\bibitem{Feng:2012iq}
F.~Feng,
\texttt{Apart}: A generalized Mathematica \texttt{Apart} function,
Comput.\ Phys.\ Commun.\ \textbf{183}, 2158 (2012).

\bibitem{Smirnov:2008iw}
A.~V.~Smirnov,
Algorithm \textsc{FIRE}---Feynman Integral REduction,
JHEP \textbf{10}, 107 (2008).

\bibitem{Hahn:1998yk}
T.~Hahn and M.~P\'erez-Victoria,
Automatized one loop calculations in four-dimensions and $D$-dimensions,
Comput.\ Phys.\ Commun.\ \textbf{118}, 153 (1999).

\bibitem{ParticleDataGroup:2026}
F.~Takahashi \textit{et al.} (Particle Data Group),
Review of Particle Physics,
Int.\ J.\ Mod.\ Phys.\ A \textbf{41}, 2630011 (2026).

\bibitem{Catani:1996vz}
S.~Catani and M.~H.~Seymour,
A general algorithm for calculating jet cross sections in NLO QCD,
Nucl.\ Phys.\ B \textbf{485}, 291 (1997); \textbf{510}, 503(E) (1998).

\end{thebibliography}
\end{document}